\documentclass[journal,final]{IEEEtran}

\usepackage[T1]{fontenc}
\usepackage{verbatim}
\usepackage{amsmath}
\usepackage{amssymb}
\usepackage{graphicx}
\usepackage{tikz}
\usetikzlibrary{positioning}
\usepackage[noadjust]{cite}
\usepackage{subcaption}
\allowdisplaybreaks

\newcommand{\partiald}[2]{\ensuremath{\frac{\partial #1}{\partial #2}}}
\definecolor{DarkBlue}{rgb}{0, 0, 0.85}
\definecolor{DarkGreen}{rgb}{0, 0.5, 0}

\newcommand\revision[1]{{#1}}
\newcommand\revisionC[1]{{#1}}
\newcommand\paperplan[2]{}
\newcommand\paperplanc[2]{}

\newcommand\togglefig[1]{#1}

\makeatletter

\providecommand{\tabularnewline}{\\}

\makeatother

\usepackage{babel}
\begin{document}
\title{Path Planning with Uncertainty for Aircraft Under Threat of Detection from Ground-Based Radar}
\author{Austin Costley, Greg Droge, Randall Christensen, Robert C. Leishman, James Swedeen
\thanks{This work was supported by Air Force Research Laboratory, Wright-Patterson
Air Force Base, OH.\protect \\
A. Costley is with the Electrical and Computer Engineering Department,
Utah State University, Logan, UT 84322 USA (e-mail: austin.costley@usu.edu)\protect \\
G. Droge is with the Electrical and Computer Engineering Department,
Utah State University, Logan, UT 84322 USA (e-mail: greg.droge@usu.edu)\protect \\
R. Christensen is with Blue Origin, Kent, WA 98032 USA (e-mail: rchristensen@blueorigin.com)\protect \\
R. Leishman is with the ANT Center, Air Force Institute of Technology,
Wright-Patterson Air Force Base, OH 45433 USA (e-mail: robert.leishman@afit.edu)\protect \\
J. Swedeen is with the Electrical and Computer Engineering Department,
Utah State University, Logan, UT 84322 USA (e-mail: james.swedeen@usu.edu)}
}

\maketitle
\begin{abstract}
Mission planners for aircraft operating under threat of detection by ground-based radar systems are concerned with the probability of detection.
Current path planning methods for such scenarios consider the aircraft pose, radar position, and radar parameters to be deterministic and known.
This paper presents a framework for incorporating uncertainty in these quantities into a radar detection model that is used by a path planner.
The \revisionC{developed} path planner evaluates the radar detection risk in the presence of uncertainties and uses linear covariance analysis to efficiently generate error budgets.
The error budgets identify the contribution of each source of uncertainty (e.g., sensor measurement noise, radar position uncertainty) to the overall variability in the probability of detection.
The framework is applied to a modified visibility graph path planner that uses the \revisionC{detection risk and its variability} to calculate path adjustments\revisionC{, which} maintain the detection risk below a specified threshold.
The results show that the framework is effective at providing actionable information to the mission planner \revisionC{that improves} the final planned path and \revisionC{reduces} the detection risk.
\end{abstract}

\section*{Nomenclature \label{sec:nomenclature}}
\addcontentsline{toc}{section}{Nomenclature}
\begin{IEEEdescription}[\IEEEusemathlabelsep\IEEEsetlabelwidth{$p_{an}$, $p_{ae}$, $p_{ad}$}]
\item[$P_D$] Approximation of probability of detection
\item[$\bar{P}_D$] Nominal value of $P_D$
\item[$\delta P_D$] Perturbation of $P_D$ about nominal
\item[$P_{fa}$] Probability of false alarm
\item[$\mathcal{S}$] Signal-to-noise ratio
\item[$\sigma_r$] Radar cross section ($m^2$)
\item[$R$] Range to target ($m$)
\item[$c_r$] \revisionC{Lumped} radar parameter \revisionC{($Jm^2/^{\circ}K$)}
\item[$\boldsymbol{x_a}$] Aircraft state vector
\item[$\boldsymbol{\bar{x}_a}$] Nominal aircraft state 
\item[$\boldsymbol{p_a^n}$] Aircraft position vector in NED frame
\item[$\boldsymbol{\Theta_a}$] Aircraft Euler angle vector                  
\item[$p_{an}$, $p_{ae}$, $p_{ad}$] Aircraft position in NED frame \revisionC{($m$)}
\item[$\phi_a$, $\theta_a$, $\psi_a$] Aircraft Euler angles roll, pitch, yaw \revisionC{($rad.$)} 
\item[$\boldsymbol{x_r}$] Radar state vector
\item[$\boldsymbol{\bar{x}_r}$] Nominal radar state 
\item[$\boldsymbol{p_r^n}$] Radar position vector in NED frame
\item[$p_{rn}$, $p_{re}$, $p_{rd}$] Radar position elements in NED frame \revisionC{($m$)}
\item[$a$, $b$, $c$] Ellipsoid RCS parameters
\item[$\alpha$] RCS azimuth angle \revisionC{($rad.$)}
\item[$\phi$] RCS elevation angle \revisionC{($rad.$)}
\item[$C_{aa}$] Aircraft pose covariance
\item[$C_{rr}$] Radar state covariance 
\item[$A_{Pa}$] Jacobian of $P_D$ w.r.t. $\boldsymbol{x_a}$ 
\item[$A_{Pr}$] Jacobian of $P_D$ w.r.t. $\boldsymbol{x_r}$ 
\item[$\sigma_{pd}$] Standard deviation of $P_D$
\item[$\boldsymbol{v^n}$] Velocity vector in NED frame
\item[$q_b^n$] Aircraft attitude quaternion
\item[$\boldsymbol{b_a^b}$] Accelerometer bias in body frame 
\item[$\boldsymbol{b_g^b}$] Gyro bias in body frame
\item[$\boldsymbol{\nu^b}$] Specific force vector in body frame   
\item[$\boldsymbol{\omega^b}$] Angular rate vector in body frame   
\item[$\boldsymbol{g^n}$] Gravity vector in NED frame
\item[$\boldsymbol{w_a}$] FOGM driving noise for accelerometer bias 
\item[$\boldsymbol{w_g}$] FOGM driving noise for gyro bias
\item[$\boldsymbol{n_\nu}$] Accelerometer noise std. dev.
\item[$\boldsymbol{n_\omega}$] Gyro noise std. dev.
\item[$\delta \theta_b^n$] Error rotation vector
\item[$\delta \boldsymbol{x_e}$] Error state vector
\item[$R_b^n$] Rotation matrix from body to NED frame
\item[$P$] EKF covariance matrix
\item[$\sigma_n$, $\sigma_e$, $\sigma_d$] GPS noise std. dev.
\item[$\sigma_{h}$] Altimeter noise std. dev.
\item[$\sigma_{\psi}$] Heading noise std. dev.
\item[$\tau_a$, $\tau_g$] FOGM time constant for IMU biases
\item[$\sigma_{a,ss}$] Accelerometer bias steady state std. dev.  
\item[$\sigma_{g,ss}$] Gyro bias steady state std. dev.  
\item[$\kappa$] Path segment curvature \revisionC{($\frac{1}{m}$)}
\item[$m_\sigma$] Multiple of $\sigma_{pd}$ used in planner
\item[$P_{DT}$] $P_D$ threshold for planning
\item[$P_{true}$] True navigation error covariance
\item[$C_A$] Augmented system covariance
\item[$k$] Boltzmann's constant    
\end{IEEEdescription}

\section{Introduction}
Aircraft mission planners are tasked with planning paths for aircraft operating under threat of detection by ground-based radar systems.
Example missions include reconnaissance \cite{Ceccarelli_micro_uav}, radar counter-measure deployment \cite{larson_path_nodate,xiao-wei_path_2010}, and combat operations \cite{kabamba_optimal_2012}.
Mission planners for such scenarios are primarily interested in the probability of being detected by a radar system.
This paper develops and demonstrates a framework that allows for the consideration of aircraft and radar state uncertainties when planning a path constrained to stay below a given probability of detection threshold.

The work herein builds upon two groups in the literature. The first is the target detection literature in which high-fidelity radar detection models have been developed \cite{marcum_statistical_1960,swerling_probability_1954,mahafza_matlab_2003}. 
In particular, \cite{mahafza_matlab_2003} defines a value, $P_D$, that approximates the probability of detecting an aircraft given the signal-to-noise ratio and false alarm rate of the detection. 
The signal-to-noise ratio is dependent upon radar characteristics, relative positioning, and the radar cross section, which can vary significantly based upon the orientation of the aircraft. The second group is the \revisionC{radar detection path} planning literature \cite{mcfarland_motion_1999,bortoff_path_2000,chandler_uav_2000,pachter_optimal_2001,moore_radar_2002,jun_path_2003,larson_path_nodate,kabamba_optimal_2012,xiao-wei_path_2010}, which use some form of detection risk that may only consider some aspects of the high-fidelity radar detection models. 
The planning literature seeks to rapidly evaluate candidate paths while the target detection literature seeks to create a high-fidelity determination of detection probabilities. Neither body considers the uncertainty of the aircraft state, radar position, or radar parameters, which are all estimated and include some uncertainty.

When accurate position measurements are available, neglecting aircraft state uncertainty is a useful technique to reduce complexity. 
However, scenarios where an aircraft is seeking to avoid radar detection may include regions where accurate position measurements are not available (i.e. GPS-denied regions).
In such regions, the state uncertainty grows and becomes a significant factor in the variability of the predicted $P_D$ \cite{costley2022sensitivity}. 
Aircraft operating in radar detection regions are often equipped with an aided inertial navigation system (INS).
Such systems use measurements from an inertial measurement unit (IMU) and aiding sources, such as GPS, to estimate the aircraft state and state covariance \cite{savage_strapdown_2000,grewal_global_2020,farrell_aided_2008}.
A common filtering technique for an aided INS is the Extended Kalman Filter (EKF) \cite{maybeck_stochastic_1994,brown_kalman}, which propagates and updates the state estimate in the presence of measurement and process noise.

There are a couple of key questions \revisionC{that must be answered} when creating paths that respect thresholds on $P_D$.
First, how does the planner calculate an estimate for $P_D$ and its uncertainty along a path while considering uncertainty in the aircraft state, radar position, and radar parameters? 
Answering this question aids in determining whether a path is valid.
Second, how does one determine which noise sources (e.g.,  measurement, process, uncertain initial conditions, etc.) are the main cause of growth in $P_D$ and its associated uncertainty (i.e., error budget analysis \cite{maybeck_stochastic_1994,farrell_aided_2008})? 
\revision{Answering this question aids in determining, for example, whether a higher quality sensor will sufficiently reduce $P_D$ for a given path.}
% This question can be answered using error budget analysis, which can help determine, for example, the contribution of the IMU measurement noise to the overall uncertainty in $P_D$.
% If the IMU measurement noise is the main cause of uncertainty in $P_D$, then a higher quality IMU sensor may result in a sufficient reduction of $P_D$ to take a given path.

A Monte Carlo approach could be utilized to answer the questions posed in the previous paragraph to an arbitrary level of fidelity. In a Monte Carlo approach, each path under consideration is simulated hundreds or thousands of times to model the effect of uncertainties on the aircraft state as estimated by the navigation filter. Each simulation, or run, uses a different sampling of the noise and radar uncertainties. 
The resulting ensemble statistics are calculated for $P_D$ at each point in time to quantify the variability in $P_D$ due to the uncertainties present in the simulations.
The Monte Carlo analysis is repeated several times with \revisionC{a different set of active error sources} to obtain the necessary data to build an error budget.  
This approach is computationally intensive and not well-suited to rapid planning or error budget analysis.

An efficient alternative to Monte Carlo analysis is linear covariance analysis \cite{maybeck_stochastic_1994,christensen_2014,christensen2021closedloop}. 
Linear covariance analysis utilizes similar linearization techniques and Gaussian noise assumptions as employed by an EKF to propagate the second order \revisionC{moments} of the random variables in question. 
This approach approximates the same statistical information as Monte Carlo analysis in a single simulation over the trajectory.

The path planner in this paper evaluates $P_D$ with respect to the covariance of the aircraft and radar states using a framework provided in \cite{costley2022sensitivity,rds_ext}.
The IMU measurement generation capability developed in \cite{asg} provides inputs to an inertial navigation filter to enable rapid evaluation of the aircraft state covariance along candidate paths. 
This paper uses linear covariance techniques to rapidly evaluate the \revisionC{variability} of $P_D$ \revisionC{due to uncertainties estimated by a navigation filter} and demonstrates the use of these statistics in a path planning application.
Given a path to be followed, the \revisionC{efficient} IMU signal generator in \cite{asg} is used to generate representative measurements along the trajectory that are incorporated into a navigation filter.
The aircraft and radar state uncertainties are used as in \cite{costley2022sensitivity,rds_ext} to estimate the variance of $P_D$. The path planner generates multiple such candidate paths and eliminates paths that violate threshold constraints.

The contributions of this work are threefold. First, a framework is developed for the calculation of the variance of $P_D$ given a trajectory. The variance incorporates the aircraft state and radar state uncertainties. Second, an error budget analysis for the resulting variance of $P_D$ is developed using linear covariance analysis. Third, an iterative path planning technique based upon shortest path visibility graphs is created to demonstrate the ability to consider $P_D$ uncertainty when planning.

The remainder of this paper is organized as follows.
Section \ref{sec:radar_model} presents the radar detection model which provides an expression for the probability of detection, the linearization of the model, and the incorporation of aircraft and radar state uncertainties.
The aircraft state uncertainty is computed using an aided-INS as described in Section \ref{sec:ins_model}.
Section \ref{sec:lincov} describes the linear covariance model and the method for generating error budgets.
Section \ref{sec:vgraph} describes the path planner that incorporates these components and the results of the path planner are provided in Section \ref{sec:Results}.

\section{Background and Previous Work \label{sec:background}}
\begin{figure}
    \begin{centering}
    \tikzstyle{startstop} = [rectangle, minimum width=2cm, minimum height=1cm,text centered, draw=black]
    \tikzstyle{process} = [rectangle, minimum width=1.9cm,  text width=1.9cm, minimum height=1cm, text centered, draw=black]
    \tikzstyle{arrow} = [thick,->,>=stealth]
    % \definecolor{bcolor}{rgb}{0.16, 0.32, 0.75}
    % \definecolor{bcolor}{rgb}{0.11, 0.22, 0.73}
    \definecolor{bcolor}{rgb}{0.0, 0.0, 0.0}
    \scalebox{0.7}{
    \begin{tikzpicture}[node distance=3.7cm]
        % Nodes
        % \node (in) [text width=1.5cm, text centered, color=bcolor] {waypoints};
        \node (in) [color=bcolor] {};
        \node (pro2) [process, color=bcolor, right of=in, xshift=-1.5cm] {ASG \\ Section \ref{sec:ASG}};
        \node (ins) [process, color=bcolor, right of=pro2] {INS \\ Section \ref{sec:ins_model}};
        \node (pro3) [process, color=bcolor, right of=ins] {Radar Model \\ Section \ref{sec:radar_model}};
        % \node (out) [text width=1.5cm, color=bcolor, text centered, right of=pro3, xshift=-1.5cm] {$\bar{P}^i_D$, $\sigma^i_{pd}$};
        \node (out) [color=bcolor, right of=pro3, xshift=-1.6cm] {};

        % Arrows
        \draw [arrow, color=bcolor, text width=1.25cm, align=center] (pro2) -- node[anchor=south] {$\boldsymbol{\bar{x}_a}$, $\boldsymbol{\nu^b}$, $\boldsymbol{\omega^b}$} (ins);
        \draw [arrow, color=bcolor] (ins) -- node[anchor=south] {$\boldsymbol{\bar{x}_a}$, $C_{aa}$} (pro3);
        \draw [arrow, color=bcolor] (in) -- node[anchor=south, near start]{waypoints} (pro2);
        \draw [arrow, color=bcolor] (pro3) -- node[anchor=south, near end]{$\bar{P}_D$, $\sigma_{pd}$}(out);
    \end{tikzpicture}
    }
    \par\end{centering}
    \caption{Block diagram for components discussed in the background section.\label{fig:background_block_diagram}}
\end{figure}
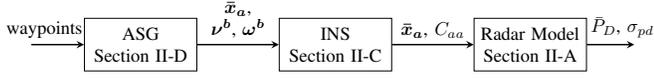

\paperplan{Background}
{
    \begin{itemize}
        \item Introduce major components
        \begin{itemize}
            \item Radar detection
            \item ASG
            \item Inertial Navigation
            \item others?
        \end{itemize}
        \item Provide necessary development and relevant background
        \item Possibly add a diagram
    \end{itemize}
}
This section describes previous work that is relevant to the path planning application presented in this paper.
The path planner seeks to keep $P_D$ below a specified threshold while accounting for uncertainty in the aircraft and radar states.
Section \ref{sec:radar_model} describes the development of the radar detection model from \cite{mahafza_matlab_2003} and the linearization of that model to obtain an expression for the variance of $P_D$ as shown in \cite{costley2022sensitivity,rds_ext}.
Section \ref{sec:ins_model} details the INS model used in this paper to calculate the aircraft pose covariance.
Finally, Section \ref{sec:ASG} describes \revision{the rapid aircraft state generation (ASG) used to create representative} aircraft state and IMU measurements from smoothed piece-wise linear paths as presented in \cite{asg}.
These measurements are used to propagate the aircraft state covariance.
The block diagram in Fig. \ref{fig:background_block_diagram} shows the relationship of the \revision{ASG, INS, and radar model} as described in this section.

%
% RADAR DETECTION SUBSECTION
\subsection{Probability of Detection \label{sec:radar_model}}
\paperplan{Radar Models}
{
    \begin{itemize}
        \item Introduce radar model
        \item Reference previous work to obtain $\sigma_{pd}$
    \end{itemize}
}
The radar detection model used in this paper is presented in \cite{mahafza_matlab_2003}.
For sake of completeness, the main equations are provided in this section.
The model includes expressions for the probability of detection, $P_D$, signal-to-noise ratio, $\mathcal{S}$, and RCS, $\sigma_r$. 
These quantities are functions of the aircraft pose, radar position, and radar parameters. 

% \subsection{Probability of Detection}
An accurate approximation to $P_D$, provided by North \cite{mahafza_matlab_2003, north_analysis_1963}, is
\begin{equation}
P_{D}\approx 0.5 \times \text{erfc}\left(\sqrt{-\ln P_{fa}}-\sqrt{\mathcal{S}+0.5}\right)\label{eq:pd_approx}
\end{equation}
where $P_{fa}$ is the probability of false alarm, $\mathcal{S}$ is the signal-to-noise ratio, and $\text{erfc}(\cdot)$
is the complementary error function.
The $P_{fa}$ is considered a constant for a given radar, whereas $\mathcal{S}$ is a function of radar parameters and the pose of the target aircraft relative to the radar.

A general expression for the signal-to-noise ratio is given by 
\begin{eqnarray}
\mathcal{S}&=&c_r\frac{\sigma_r}{kR^4}\label{eq:SNR}
\end{eqnarray}
where $k$ is Boltzmann's constant $(1.38\times10^{-23} \; J/^{\circ}K)$ and $c_r$ is a radar constant that is a function of several radar parameters. 
These parameters include power, aperture area, noise figure, and loss factor.
The specific equation for $c_r$ is dependent on the type of radar being modeled \cite{mahafza_matlab_2003}.

\begin{figure}
    \centering
    \tikzset{every picture/.style={line width=0.75pt}} %set default line width to 0.75pt        

\begin{tikzpicture}[x=0.75pt,y=0.75pt,yscale=-1,xscale=1]

%Straight Lines [id:da3425513975196144] 
\draw    (146,241) -- (146,133) ;
\draw [shift={(146,131)}, rotate = 90] [color={rgb, 255:red, 0; green, 0; blue, 0 }  ][line width=0.75]    (10.93,-3.29) .. controls (6.95,-1.4) and (3.31,-0.3) .. (0,0) .. controls (3.31,0.3) and (6.95,1.4) .. (10.93,3.29)   ;
%Straight Lines [id:da7387825232985139] 
\draw    (146,241) -- (264,241) ;
\draw [shift={(266,241)}, rotate = 180] [color={rgb, 255:red, 0; green, 0; blue, 0 }  ][line width=0.75]    (10.93,-3.29) .. controls (6.95,-1.4) and (3.31,-0.3) .. (0,0) .. controls (3.31,0.3) and (6.95,1.4) .. (10.93,3.29)   ;
%Shape: Rectangle [id:dp657006035048564] 
\draw  [color={rgb, 255:red, 128; green, 128; blue, 128 }  ,draw opacity=1 ][fill={rgb, 255:red, 128; green, 128; blue, 128 }  ,fill opacity=1 ] (263.17,105.53) -- (320.08,86.24) -- (323.29,95.71) -- (266.38,115.01) -- cycle ;
%Shape: Circle [id:dp05377977243413867] 
\draw  [color={rgb, 255:red, 128; green, 128; blue, 128 }  ,draw opacity=1 ][fill={rgb, 255:red, 128; green, 128; blue, 128 }  ,fill opacity=1 ] (316.95,92.58) .. controls (316.06,89.96) and (317.46,87.13) .. (320.08,86.24) .. controls (322.69,85.35) and (325.53,86.75) .. (326.42,89.37) .. controls (327.31,91.98) and (325.91,94.82) .. (323.29,95.71) .. controls (320.68,96.6) and (317.84,95.2) .. (316.95,92.58) -- cycle ;
%Shape: Half Frame [id:dp8259964541100322] 
\draw  [color={rgb, 255:red, 128; green, 128; blue, 128 }  ,draw opacity=1 ][fill={rgb, 255:red, 128; green, 128; blue, 128 }  ,fill opacity=1 ] (308.1,95.19) -- (291.93,132.17) -- (288.41,122.93) -- (298.68,99.45) -- (275.63,89.38) -- (271.81,79.32) -- cycle ;
%Shape: Diagonal Stripe [id:dp6793600857980076] 
\draw  [color={rgb, 255:red, 128; green, 128; blue, 128 }  ,draw opacity=1 ][fill={rgb, 255:red, 128; green, 128; blue, 128 }  ,fill opacity=1 ] (276.17,109.95) -- (268.87,123.66) -- (259.73,96.14) -- (274.13,103.8) -- cycle ;
%Straight Lines [id:da06808538152572452] 
\draw  [dash pattern={on 4.5pt off 4.5pt}]  (293.23,100.62) -- (293.23,62.62) ;
\draw [shift={(293.23,60.62)}, rotate = 90] [color={rgb, 255:red, 0; green, 0; blue, 0 }  ][line width=0.75]    (10.93,-3.29) .. controls (6.95,-1.4) and (3.31,-0.3) .. (0,0) .. controls (3.31,0.3) and (6.95,1.4) .. (10.93,3.29)   ;
%Straight Lines [id:da6634380532555486] 
\draw  [dash pattern={on 4.5pt off 4.5pt}]  (293.23,100.62) -- (331.23,100.62) ;
\draw [shift={(333.23,100.62)}, rotate = 180] [color={rgb, 255:red, 0; green, 0; blue, 0 }  ][line width=0.75]    (10.93,-3.29) .. controls (6.95,-1.4) and (3.31,-0.3) .. (0,0) .. controls (3.31,0.3) and (6.95,1.4) .. (10.93,3.29)   ;
%Straight Lines [id:da4779799555380675] 
\draw    (293.23,100.62) -- (323.14,89.93) -- (362.12,76.01) ;
\draw [shift={(364,75.33)}, rotate = 160.34] [color={rgb, 255:red, 0; green, 0; blue, 0 }  ][line width=0.75]    (10.93,-3.29) .. controls (6.95,-1.4) and (3.31,-0.3) .. (0,0) .. controls (3.31,0.3) and (6.95,1.4) .. (10.93,3.29)   ;
%Shape: Arc [id:dp8818629584090831] 
\draw  [draw opacity=0] (293.99,83.13) .. controls (301.4,83.46) and (307.58,88.68) .. (309.57,95.73) -- (293.23,100.62) -- cycle ; \draw   (293.99,83.13) .. controls (301.4,83.46) and (307.58,88.68) .. (309.57,95.73) ;
%Straight Lines [id:da7624498539857514] 
\draw    (293.23,100.62) -- (200,200) ;
%Shape: Arc [id:dp8397165372919599] 
\draw  [draw opacity=0] (199.61,180) .. controls (199.74,180) and (199.87,180) .. (200,180) .. controls (205.39,180) and (210.46,181.42) .. (214.83,183.92) -- (200,210) -- cycle ; \draw   (199.61,180) .. controls (199.74,180) and (199.87,180) .. (200,180) .. controls (205.39,180) and (210.46,181.42) .. (214.83,183.92) ;
%Shape: Arc [id:dp3715691837693851] 
\draw  [draw opacity=0] (342.2,83.06) .. controls (344.25,88.54) and (345.37,94.45) .. (345.37,100.62) .. controls (345.37,128.85) and (322.02,151.73) .. (293.23,151.73) .. controls (279.68,151.73) and (267.33,146.66) .. (258.06,138.34) -- (293.23,100.62) -- cycle ; \draw   (342.2,83.06) .. controls (344.25,88.54) and (345.37,94.45) .. (345.37,100.62) .. controls (345.37,128.85) and (322.02,151.73) .. (293.23,151.73) .. controls (279.68,151.73) and (267.33,146.66) .. (258.06,138.34) ;
%Shape: Diamond [id:dp4656154718136989] 
\draw  [color={rgb, 255:red, 0; green, 0; blue, 0 }  ,draw opacity=1 ][fill={rgb, 255:red, 0; green, 0; blue, 0 }  ,fill opacity=1 ] (200,190) -- (210,200) -- (200,210) -- (190,200) -- cycle ;
%Shape: Circle [id:dp03463800803247641] 
\draw  [fill={rgb, 255:red, 0; green, 0; blue, 0 }  ,fill opacity=1 ] (290.73,100.62) .. controls (290.73,99.24) and (291.85,98.12) .. (293.23,98.12) .. controls (294.61,98.12) and (295.73,99.24) .. (295.73,100.62) .. controls (295.73,102) and (294.61,103.12) .. (293.23,103.12) .. controls (291.85,103.12) and (290.73,102) .. (290.73,100.62) -- cycle ;
%Straight Lines [id:da4847368434141208] 
\draw  [dash pattern={on 4.5pt off 4.5pt}]  (200,200) -- (200,142) ;
\draw [shift={(200,140)}, rotate = 90] [color={rgb, 255:red, 0; green, 0; blue, 0 }  ][line width=0.75]    (10.93,-3.29) .. controls (6.95,-1.4) and (3.31,-0.3) .. (0,0) .. controls (3.31,0.3) and (6.95,1.4) .. (10.93,3.29)   ;
%Straight Lines [id:da22722452086323575] 
\draw  [dash pattern={on 4.5pt off 4.5pt}]  (200,200) -- (248,200) ;
\draw [shift={(250,200)}, rotate = 180] [color={rgb, 255:red, 0; green, 0; blue, 0 }  ][line width=0.75]    (10.93,-3.29) .. controls (6.95,-1.4) and (3.31,-0.3) .. (0,0) .. controls (3.31,0.3) and (6.95,1.4) .. (10.93,3.29)   ;
%Straight Lines [id:da4313105626564906] 
\draw    (293.23,100.62) -- (305.31,133.46) ;
\draw [shift={(306,135.33)}, rotate = 249.8] [color={rgb, 255:red, 0; green, 0; blue, 0 }  ][line width=0.75]    (10.93,-3.29) .. controls (6.95,-1.4) and (3.31,-0.3) .. (0,0) .. controls (3.31,0.3) and (6.95,1.4) .. (10.93,3.29)   ;

% Text Node
\draw (301,59.4) node [anchor=north west][inner sep=0.75pt]    {$ \begin{array}{l}
\psi _{a}\\
\end{array}$};
% Text Node
\draw (200,160.4) node [anchor=north west][inner sep=0.75pt]    {$ \begin{array}{l}
\theta _{r}\\
\end{array}$};
% Text Node
\draw (336,131.4) node [anchor=north west][inner sep=0.75pt]    {$\alpha $};
% Text Node
\draw (127,122) node [anchor=north west][inner sep=0.75pt]   [align=left] {N};
% Text Node
\draw (277,242) node [anchor=north west][inner sep=0.75pt]   [align=left] {E};
% Text Node
\draw (181,211) node [anchor=north west][inner sep=0.75pt]   [align=left] {radar};
% Text Node
\draw (227,61.4) node [anchor=north west][inner sep=0.75pt]  [font=\scriptsize]  {$\begin{bmatrix}
p_{an}\\
p_{ae}\\
p_{ad}
\end{bmatrix}$};
% Text Node
\draw (157,161.4) node [anchor=north west][inner sep=0.75pt]  [font=\scriptsize]  {$\begin{bmatrix}
p_{rn}\\
p_{re}\\
p_{rd}
\end{bmatrix}$};
% Text Node
\draw (352,59.4) node [anchor=north west][inner sep=0.75pt]  [font=\scriptsize]  {$b_{x}$};
% Text Node
\draw (311,121.4) node [anchor=north west][inner sep=0.75pt]  [font=\scriptsize]  {$b_{y}$};

\end{tikzpicture}    
\caption{ Graphical representation of the quantities used in the radar detection model.\label{fig:radar_xy}}
\end{figure}
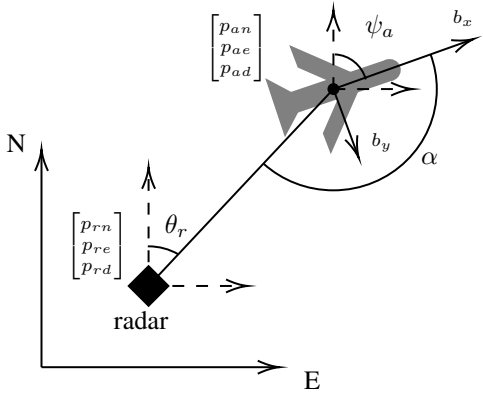

The RCS, $\sigma_{r}$, and range, $R$, are functions of the aircraft pose and radar position as depicted in Fig. \ref{fig:radar_xy}.
The aircraft pose consists of the position in the North-East-Down (NED) frame, $\boldsymbol{p_a^n}$, and a vector of the Euler angles, $\boldsymbol{\Theta_a}$, as described in \cite{beard_randy_small_2012}. 
The range to the radar is given by
\begin{eqnarray}
R & = & ||\boldsymbol{p_{a}^n}-\boldsymbol{p_{r}^n}||_{2} \label{eq:range}
\end{eqnarray}
where $\boldsymbol{p_{r}^n}$ represents the position of the radar in the NED frame.

The framework developed in \cite{costley2022sensitivity,rds_ext} supports several RCS models.
In this paper, the ellipsoid RCS representation is used \cite{mahafza_matlab_2003,kabamba_optimal_2012,costley2022sensitivity}, which is a function of the RCS azimuth $\alpha$ and elevation $\phi$ angles and is given by
\begin{align}
\sigma_{r}=\frac{\pi\left(abc\right)^{2}}
    {\left(\left(a \,\textrm{S}\alpha \,\textrm{C}\phi\right)^{2}+\left(b \,\textrm{S}\alpha\, \textrm{S}\phi\right)^{2}+\left(c\, \textrm{C}\alpha\right)^{2}\right)^{2}} \label{eq:rcs_ellipsoid}
\end{align}
where $a$, $b$, and $c$ are the length of the ellipsoid axes and $\textrm{S}\cdot$ and $\textrm{C}\cdot$ are the $\sin(\cdot)$ and $\cos(\cdot)$ functions. 
Equations for the RCS azimuth and elevation angles as functions of the aircraft pose and radar position are provided in \cite{costley2022sensitivity,rds_ext}. %Appendix \ref{app:az_el}.

\subsection{Uncertainty in Probability of Detection}
The framework presented in \cite{costley2022sensitivity,rds_ext} provides a method for incorporating aircraft and radar state uncertainty into the calculation of $P_D$ for a single-pulse radar detection model.
The framework provides an expression for the variance of $P_D$ given the covariance of the aircraft and radar states. 

Let the aircraft state be $\boldsymbol{x_a} = \begin{bmatrix} \boldsymbol{p_a^n} & \boldsymbol{\Theta_a} \end{bmatrix}^\intercal$ and radar state be $\boldsymbol{x_r} = \begin{bmatrix} \boldsymbol{p_r^n} & c_r \end{bmatrix}^\intercal$. 
The state vectors are represented as Gaussian distributed random variables with $\boldsymbol{x_a} \sim \mathcal{N}\left(\boldsymbol{\bar{x}_a}, C_{aa}\right)$ and $\boldsymbol{x_r} \sim \mathcal{N}\left(\boldsymbol{\bar{x}_r}, C_{rr}\right)$ where the bar notation is used to indicate the mean (or nominal).
\revisionC{To find a linear approximation of $P_D$, $\boldsymbol{x_a}$ and $\boldsymbol{x_r}$ are both expressed as a the sum of a nominal state and a perturbation as}
\begin{align}
    \boldsymbol{x_a} &= \boldsymbol{\bar{x}_a} + \delta \boldsymbol{x_a} \\
    \boldsymbol{x_r} &= \boldsymbol{\bar{x}_r} + \delta \boldsymbol{x_r}.
\end{align}
Because $P_D$ is a nonlinear function of the aircraft and radar state, variability in $\boldsymbol{x_a}$ and $\boldsymbol{x_r}$ induce variability in $P_D$.
Thus, $P_D$ is expressed as the sum of a nominal and a perturbation as
\begin{align}
    P_D = \bar{P}_D + \delta P_D.
\end{align}
The perturbation, $\delta P_D$, is approximated, to the first order, by linearizing \eqref{eq:pd_approx}-\eqref{eq:rcs_ellipsoid} about the nominal operating points ($\boldsymbol{\bar{x}_a}$, $\boldsymbol{\bar{x}_r}$) using a Taylor series expansion to obtain
\begin{align}
    % \delta P_{D} & \approx \partiald{P_D}{\mathcal{S}}\left(\partiald{\mathcal{S}}{R}\partiald{R}{\boldsymbol{x_{a}}}+\partiald{\mathcal{S}}{\sigma_{r}}\partiald{\sigma_{r}}{\boldsymbol{x_{a}}}\right) \Bigg\rvert_{\boldsymbol{\bar{x}_a}} \delta\boldsymbol{x_{a}} + \nonumber \\
    %              & \qquad \partiald{P_D}{\mathcal{S}}\begin{bmatrix}\partiald{\mathcal{S}}{R}\partiald{R}{\boldsymbol{p_r^n}}+\partiald{\mathcal{S}} {\sigma_{r}}\partiald{\sigma_{r}}{\boldsymbol{p_r^n}} & \partiald{\mathcal{S}}{c_r} \end{bmatrix}\Bigg\rvert_{\boldsymbol{\bar{x}_r}}\delta\boldsymbol{x_{r}} \label{eq:pd_partial_long}\\
    %  & \approx A_{Pa}\delta\boldsymbol{x_{a}} + A_{Pr}\delta\boldsymbol{x_{r}}. \label{eq:pd_partial_short}
    \delta P_{D} &\approx A_{Pa}\delta\boldsymbol{x_{a}} + A_{Pr}\delta\boldsymbol{x_{r}}. \label{eq:pd_partial_short}
\end{align}
The Jacobians $A_{Pa}$ and $A_{Pr}$ are the partial derivatives of $P_D$ as defined in \eqref{eq:pd_approx} with respect to $\boldsymbol{x_a}$ and $\boldsymbol{x_r}$, respectively.
The partial derivatives are described in detail in \cite{costley2022sensitivity,rds_ext}.% and are provided in Appendix \ref{app:rad_jacs} for reference.

The variance of $P_D$ due to aircraft and radar state covariance is computed by taking the expectation of $\delta P_D^2$ as
\begin{align}
\sigma_{pd}^2 & = E\left[\delta P_{D}^2\right]\\
  & = A_{Pa} C_{aa} A_{Pa}^\intercal + A_{Pr} C_{rr} A_{Pr}^\intercal. \label{eq:pd_var}
\end{align}
These equations show that variability of $P_D$ is a function of the Jacobians $A_{Pa}$ and $A_{Pr}$, the covariance of the aircraft state, $C_{aa}$, and the covariance of the radar state, $C_{rr}$. 
For the path planning application presented in this paper, $C_{rr}$ is considered constant and $C_{aa}$ is time-varying and \revisionC{obtained from a stochastic model of an inertial navigation system.}

%
% INS SUBSECTION
\subsection{Inertial Navigation Model\label{sec:ins_model}}
\paperplan{Inertial Navigation}
{
    \begin{itemize}
        \item Dynamic model of aircraft
        \begin{itemize}
            \item Driven by IMU measurements
            \item Quaternion attitude representation
        \end{itemize}
        \item ASG for state and IMU Generation
        \begin{itemize}
            \item Need to Convert ASG Euler angles to quaternion 
        \end{itemize}
        \item Propagate covariance using EKF $\dot{P}$ equation
        \begin{itemize}
            \item F matrix has IMU measurements
            \item Lear's method for better efficiency
        \end{itemize}
        \item Update covariance using EKF
        \begin{itemize}
            \item GPS and altitude measurements
            \item Others(?)
        \end{itemize}
        \item Results in time-varying Covariance matrix $C_{xx}$
    \end{itemize}
}

The expression in \eqref{eq:pd_var} indicates that the variance of $P_D$ is a function of the aircraft state covariance, $C_{aa}$.
This section provides a method for calculating $C_{aa}$ by modeling the covariance propagation of an aided-INS.

The INS used in this work is a continuous-time error-state EKF with discrete measurement updates for position, heading, and altitude.
The covariance of the EKF is propagated \revisionC{with} measurements from an IMU using the ``model replacement'' method \cite{maybeck_stochastic_1994}.
The following paragraphs describe the method for propagating and updating the aircraft state covariance for the INS. 

The error-state EKF estimates the difference between a truth and navigation model of the aircraft dynamics. 
For the development of this filter, let the truth model for the aircraft dynamics be defined as
\begin{equation}
    \underset{\dot{\boldsymbol{x}}}{\underbrace{\left[\begin{array}{c}
    \dot{\boldsymbol{p}}^{n}\\
    \dot{\boldsymbol{v}}^{n}\\
    \dot{q}_{b}^{n}\\
    \dot{\boldsymbol{b}}^b_{a}\\
    \dot{\boldsymbol{b}}^b_{g}
    \end{array}\right]}}=\underset{\boldsymbol{f}\left(\boldsymbol{x},\boldsymbol{w},t\right)}{\underbrace{\left[\begin{array}{c}
    \boldsymbol{v}^{n}\\
    T_{b}^{n}\boldsymbol{\nu}^{b}\left(t\right)+\boldsymbol{g}^{n}\\
    \frac{1}{2}q_{b}^{n}\otimes\left[\begin{array}{c}
    0\\
    \boldsymbol{\omega}^{b}\left(t\right)
    \end{array}\right]\\
    -\frac{1}{\tau_{a}}\boldsymbol{b}^b_{a}+\boldsymbol{w}_{a}\\
    -\frac{1}{\tau_{g}}\boldsymbol{b}^b_{g}+\boldsymbol{w}_{g}
    \end{array}\right]}}\label{eq:truth_model}
\end{equation}
where the aircraft position, $\boldsymbol{p}^{n}$, and
velocity, $\boldsymbol{v}^{n}$, are in the NED frame,
the attitude quaternion, $q_{b}^{n}$, is the orientation of the body frame with respect to the NED frame, and the accelerometer bias $\boldsymbol{b}_{a}^{b}$,
and gyro bias $\boldsymbol{b}_{g}^{b}$ are in the body
frame.
The matrix $T^n_b$ is the \revisionC{transformation} from the body frame to the NED frame associated with the attitude quaternion $q_b^n$ and $\otimes$ represents the Hamiltonian quaternion product \cite{zanetti_rotations_2019}.
The accelerometer and gyro biases are modeled as Fist-Order Gauss Markov (FOGM) processes with time constants $\tau_a$ and $\tau_g$ and driving white noise of $\boldsymbol{w}_a$ and $\boldsymbol{w}_g$.
The truth model is driven by the true specific force $\boldsymbol{\nu}^b$ and angular rate $\boldsymbol{\omega}^b$ of the aircraft body.
This model is similar to navigation models presented in \cite{savage_strapdown_2000,grewal_global_2020,farrell_aided_2008}
with the addition of FOGM sensor biases as in \cite{maybeck_stochastic_1994}.

The navigation model has the same states as the truth model but is driven by biased and noisy measurements from an IMU with
\begin{equation}
    \underset{\dot{\hat{\boldsymbol{x}}}}{\underbrace{\left[\begin{array}{c}
    \dot{\hat{\boldsymbol{p}}}^{n}\\
    \dot{\hat{\boldsymbol{v}}}^{n}\\
    \dot{\hat{q}}_{b}^{n}\\
    \dot{\hat{\boldsymbol{b}}}_{a}^{b}\\
    \dot{\hat{\boldsymbol{b}}}_{g}^{b}
    \end{array}\right]}}=\underset{\boldsymbol{\hat{f}}\left(\hat{\boldsymbol{x}},\tilde{\boldsymbol{y}},t\right)}{\underbrace{\left[\begin{array}{c}
    \hat{\boldsymbol{v}}^{n}\\
    \hat{T}_{b}^{n}\left(\tilde{\boldsymbol{\nu}}^{b}(t)-\hat{\boldsymbol{b}}_{a}\right)+\boldsymbol{g}^{n}\\
    \frac{1}{2}\hat{q}_{b}^{n}\otimes\left[\begin{array}{c}
    0\\
    \tilde{\boldsymbol{\omega}}^{b}(t)-\hat{\boldsymbol{b}}_{g}^{b}
    \end{array}\right]\\
    -\frac{1}{\tau_{a}}\hat{\boldsymbol{b}}_{a}^{b}\\
    -\frac{1}{\tau_{g}}\hat{\boldsymbol{b}}_{g}^{b}
    \end{array}\right]}}.\label{eq:nav_model}
\end{equation}
where the hat symbol ($\hat{\boldsymbol{x}}$) is used to indicate navigation model states.
The IMU measurements are provided by a three-axis accelerometer and gyro and are corrupted by bias and noise given by
\begin{equation}
    \underset{\boldsymbol{\tilde{y}}}{\underbrace{\left[\begin{array}{c}
    \tilde{\boldsymbol{\nu}}^{b}(t)\\
    \tilde{\boldsymbol{\omega}}^{b}(t)
    \end{array}\right]}}=
    \underset{\boldsymbol{c}(\boldsymbol{x}, \boldsymbol{u})}{\underbrace{\left[\begin{array}{c}
    \boldsymbol{\nu}^{b}(t)+\boldsymbol{b}_{a}^{b}\\
    \boldsymbol{\omega}^{b}(t)+\boldsymbol{b}_{g}^{b}
    \end{array}\right]}}
    +
    \underset{\boldsymbol{\eta}}{\underbrace{\left[\begin{array}{c}
    \boldsymbol{n}_{\nu}\\
    \boldsymbol{n}_{\omega}
    \end{array}\right]}}.\label{eq:cov-cont-inertial-measurements}
\end{equation}
where $\boldsymbol{n}_{\nu} \sim \mathcal{N}(0, Q_\nu)$ and $\boldsymbol{n}_{\omega} \sim \mathcal{N}(0, Q_\omega)$.

The EKF states are updated by discrete-time measurements using the Kalman update equation
\begin{equation}
    \hat{\boldsymbol{x}}_{k}^{+}=\hat{\boldsymbol{x}}_{k}^{-}+K_{k}\left[\tilde{\boldsymbol{z}}_{k}-\hat{\tilde{\boldsymbol{z}}}_{k}\right]\label{eq:ekf_update}
\end{equation}
where $K_k$ is the Kalman gain. The measurements are generated with
\begin{equation}
    \tilde{\boldsymbol{z}}_k = \boldsymbol{h}(\boldsymbol{x}_k) + \boldsymbol{\nu}_k
\end{equation}
and the expected measurement is given by
\begin{equation}
    \hat{\tilde{\boldsymbol{z}}}_{k}=\hat{\boldsymbol{h}}\left(\hat{\boldsymbol{x}}_{k}\right).\label{eq:nonlin7}
\end{equation}
The INS in this paper processes discrete-time measurements for position, heading, and altitude corrupted by additive white noise as
\begin{equation}
    \tilde{\boldsymbol{p}}^{n}\left[t_{k}\right]=\boldsymbol{p}^{n}\left[t_k\right]+\boldsymbol{n}_{p}\left[t_{k}\right]\label{eq:veh5}
\end{equation}
\begin{equation}
    \tilde{\psi}\left[t_{k}\right]=\psi\left[t_{k}\right]+n_{\psi}\left[t_{k}\right]\label{eq:veh5-2}
\end{equation}
\begin{equation}
    \tilde{h}\left[t_{k}\right]=h\left[t_{k}\right]+n_{h}\left[t_{k}\right]\label{eq:veh5-1}
\end{equation}
where
\begin{equation}
    \boldsymbol{n}_{p} \sim \mathcal{N}\left(\boldsymbol{0}_{3\times1},R_p=\textrm{diag}\left(\sigma_{n}^{2}, \; \sigma_{e}^{2}, \; \sigma_{d}^{2} \right)\right)
\end{equation}
\begin{equation}
    n_{\psi} \sim \mathcal{N}\left(0,R_{\psi}=\sigma_{\psi}^{2}\right)
\end{equation}
and
\begin{equation}
    n_{h} \sim \mathcal{N}\left(0,R_{h}=\sigma_{h}^{2}\right).
\end{equation}

The error states of the EKF are defined as the difference between the true and navigation states given by
\begin{equation}
    \delta \boldsymbol{x}_e = \boldsymbol{x} - \hat{\boldsymbol{x}}.
\end{equation}
where the difference is defined by subtraction for all but the attitude states.
The attitude difference is defined using quaternion arithmetic as
\begin{equation}
    \left[\begin{array}{c}
    1\\
    -\frac{1}{2}\delta\boldsymbol{\theta}{}_{b}^{n}
    \end{array}\right]=q_{b}^{n}\otimes\left(\hat{q}_{b}^{n}\right)^{*} \label{eq:quat_error}
\end{equation}
where $\left(\hat{q}_{b}^{n}\right)^{*}$ represents the quaternion conjugate of the navigation attitude quaternion and $\delta\boldsymbol{\theta}{}_{b}^{n}$ is the error rotation vector.
The error state vector is then given by
\begin{equation}
    \delta \boldsymbol{x}_e = \begin{bmatrix} \delta \boldsymbol{p}^n & \delta \boldsymbol{v}^n & \delta \boldsymbol{\theta}_b^n & \delta \boldsymbol{b}_a^b & \delta \boldsymbol{b}_g^b \end{bmatrix}^\intercal. \label{eq:error_states}
\end{equation}
    
The error state dynamics are linearized about the nominal aircraft trajectory to obtain  
\begin{equation}
    \delta\dot{\boldsymbol{x}}_e=\hat{F}\delta\boldsymbol{x}+\hat{B}\boldsymbol{w_e} \label{eq:cov-err-state-diffeq}
\end{equation}
where $\hat{F}$ is the linearized error state dynamics matrix given by
\begin{equation}
    \hat{F}=\left[\begin{array}{ccccc}
    \boldsymbol{0} & \boldsymbol{I} & \boldsymbol{0} & \boldsymbol{0} & \boldsymbol{0}\\
    \boldsymbol{0} & \boldsymbol{0} & \left[\hat{T}_{b}^{n}\left(\tilde{\boldsymbol{\nu}}^{b}-\hat{\boldsymbol{b}}_{a}^{b}\right)\right]\times & -\hat{T}_{b}^{n} & \boldsymbol{0}\\
    \boldsymbol{0} & \boldsymbol{0} & \boldsymbol{0} & \boldsymbol{0} & \hat{T}_{b}^{n}\\
    \boldsymbol{0} & \boldsymbol{0} & \boldsymbol{0} & -\frac{1}{\tau_{a}}\boldsymbol{I} & \boldsymbol{0}\\
    \boldsymbol{0} & \boldsymbol{0} & \boldsymbol{0} & \boldsymbol{0} & -\frac{1}{\tau_{g}}\boldsymbol{I}
    \end{array}\right]\label{eq:cov-F}
    \end{equation}
and $\hat{B}$ is the noise mixing matrix given by
\begin{equation}
    \hat{B}=\left[\begin{array}{cccc}
    \boldsymbol{0} & \boldsymbol{0} & \boldsymbol{0} & \boldsymbol{0}\\
    -\hat{T}_{b}^{n} & \boldsymbol{0} & \boldsymbol{0} & \boldsymbol{0}\\
    \boldsymbol{0} & \hat{T}_{b}^{n} & \boldsymbol{0} & \boldsymbol{0}\\
    \boldsymbol{0} & \boldsymbol{0} & \boldsymbol{I} & \boldsymbol{0}\\
    \boldsymbol{0} & \boldsymbol{0} & \boldsymbol{0} & \boldsymbol{I}
    \end{array}\right]. \label{eq:cov-B}
\end{equation}
The boldface $\boldsymbol{0}$ and $\boldsymbol{I}$ entries represent $3 \times 3$ zero matrix and identity matrix, respectively.  
The additive white noise vector, $\boldsymbol{w_e}$, consists of the accelerometer and
gyro measurement noise as well as the process noise of the FOGM biases
\begin{equation}
\boldsymbol{w_e}=\left[\begin{array}{cccc}
\boldsymbol{n}_{\nu} & \boldsymbol{n}_{\omega} & \boldsymbol{w}_{a} & \boldsymbol{w}_{g}\end{array}\right]^{T} \label{eq:we}
\end{equation}
which has power spectral density, $Q$, defined by
\begin{equation}
E\left[\boldsymbol{w_e}\left(t\right)\boldsymbol{w_e}\left(t'\right)^{T}\right]=Q\delta\left(t-t'\right) \label{eq:Q_PSD}
\end{equation}
where
\begin{equation}
    Q=\left[\begin{array}{cccc}
    Q_{\nu} & \boldsymbol{0} & \boldsymbol{0} & \boldsymbol{0}\\
    \boldsymbol{0} & Q_{\omega} & \boldsymbol{0} & \boldsymbol{0}\\
    \boldsymbol{0} & \boldsymbol{0} & q_{a}\boldsymbol{I} & \boldsymbol{0}\\
    \boldsymbol{0} & \boldsymbol{0} & \boldsymbol{0} & q_{g}\boldsymbol{I}
    \end{array}\right] \label{eq:PSD_Q}
\end{equation}
and
\begin{align}
    q_{a}&=\frac{2\sigma_{a,ss}^{2}}{\tau_{a}} \label{eq:qa} \\
    q_{g}&=\frac{2\sigma_{g,ss}^{2}}{\tau_{g}}. \label{eq:qg}
\end{align}

The state covariance, $P$, is propagated with \revisionC{the continuous Ricatti equation given by} 
\begin{equation}
    \dot{P} = \hat{F}P + P\hat{F}^\intercal + \hat{B}Q\hat{B}^\intercal. \label{eq:ekf_covprop}
\end{equation}
\revisionC{In implementation, the continuous Ricatti equation is often replaced by a more efficient discrete-time propagation equation that uses the state transition matrix \cite{carpenter_navigation_2018}.}
The path planner presented in this paper uses such a method where the state transition matrix is approximated using Lear's method \cite{lear_kalman_1985}.
More details about Lear's method and this covariance propagation approach are provided in Appendix \ref{app:lears}.

The EKF state covariance is updated at \revisionC{the discrete measurement times using the Joseph form} \cite{maybeck_stochastic_1994} as
\begin{equation}
P_{k}^{+}=\left(I-K_{k}H_{k}\right)P_{k}^{-}\left(I-K_{k}H_{k}\right)^{T}+K_{k}R_{\nu}K_{k}^{T}\label{eq:ekf_cov_update}
\end{equation}
where $K_k$ and $H_k$ are the Kalman gain and the measurement model Jacobian for the measurement being processed.
The measurement model Jacobian for the position, altitude, and heading measurements are given by
\begin{equation}
H_{p}=\left[\begin{array}{ccccc}
I_{3\times3} & 0_{3\times3} & 0_{3\times3} & 0_{3\times3} & 0_{3\times3}\end{array}\right] \label{eq:ekf_p_meas}
\end{equation}
\begin{equation}
H_{h}=\left[\begin{array}{ccccccc}
0 & 0 & -1 & 0_{1\times3} & 0_{1\times3} & 0_{1\times3} & 0_{1\times3}\end{array}\right] \label{eq:ekf_h_meas}
\end{equation}
and
\begin{equation}
H_{\psi}=\left[\begin{array}{ccccccc}
0_{1\times3} & 0_{1\times3} & 0 & 0 & -1 & 0_{1\times3} & 0_{1\times3}\end{array}\right]. \label{eq:ekf_psi_meas}
\end{equation}
The Kalman gain, $K_{k}$, is calculated with
\begin{equation}
K_{k}=P_{k}^{-}H_{k}\left(H_{k}P_{k}^{-}H_{k}^{T}+R_{\nu}\right)^{-1}.\label{eq:nonlin8}
\end{equation}

The INS state vector defined in this section differs from the aircraft state vector used in the radar detection model in Section \ref{sec:radar_model}.
The primary difference is that $\boldsymbol{x_a}$ is a reduced set of states and the attitude is represented by a vector of Euler angles instead of a quaternion.
The aircraft state covariance, $C_{aa}$, used in the radar detection model is a transformation of the EKF state covariance estimated by the navigation model given by
\begin{equation}
    C_{aa}(t) = M_a P(t) M_a^\intercal
\end{equation}
where $M_a$ is derived in Appendix \ref{app:quat2eul}.

%
% ASG SUBSECTION
\subsection{Aircraft State and IMU Generation \label{sec:ASG}}
\paperplan{ASG}{
    \begin{itemize}
        \item Discuss ASG
        \item Show generation of nominal states
        \item Discuss need for IMU measurements
    \end{itemize}
}

\revisionC{The previous section shows that the aircraft state covariance, $C_{aa}$, is propagated using the aircraft states, process noise and bias parameters, and measured accelerations from an IMU (i.e., $\boldsymbol{\tilde{\nu}^b}$ in \eqref{eq:cov-F}).}
\revisionC{The process noise and bias parameters are constant for a given scenario, however} the aircraft states and IMU measurements must be generated for each candidate path considered by the path planner.
To enable rapid planning, an efficient aircraft state and IMU measurement generator is desired.
This paper uses the ASG method presented in \cite{asg} to accomplish this task.

The ASG method converts a series of 2D waypoints to a smooth flyable trajectory constrained by maximum curvature and maximum curvature rate.
The waypoint path is smoothed using fillets with line, arc, and clothoid \cite{scheuer_continuous-curvature_1997} segments.
The nominal aircraft position, heading, and curvature are obtained using the path segment geometry equations.
For example, the clothoid segment geometry is defined by
\begin{eqnarray}
x(s) &=& x_{0}+\int_{0}^{s}\cos(0.5\kappa^\prime\xi^{2}+\kappa_{0}\xi+\psi_{0})d\xi\label{eq:fresnel_x} \\
y(s) &=& y_{0}+\int_{0}^{s}\sin(0.5\kappa^\prime\xi^{2}+\kappa_{0}\xi+\psi_{0})d\xi\label{eq:fresnel_y} \\
\psi(s) &=& \psi_{0}+\kappa_{0}s+0.5\kappa^\prime s^{2} \label{eq:clothoid_heading} \\
\kappa(s) &=& \kappa_{0}+\kappa^\prime s \label{1eq:clothoid_curvature}
\end{eqnarray}
where $s$ is the length along the segment, $\psi_{0}$ is the initial heading, $\kappa_{0}$ is the initial curvature, $x_{0}$ and $y_{0}$ represent the starting point, and $\kappa^\prime$ is the curvature rate per unit length of the segment.
This approach enables efficient generation of nominal aircraft states that adhere to a flyable trajectory given the vehicle maneuver constraints.

In the ASG method, the pitch angle, $\theta_a$, is a constant trim value and the roll angle, $\phi_a$, is obtained from a coordinated turn model \cite{beard_randy_small_2012}.
The coordinated turn model provides a relationship between the heading rate and roll angle of the aircraft as
\begin{equation}
    \dot{\psi_a}=\frac{g}{\dot{s}}\tan\phi_a
    \label{eq:coordinated_turn_roll_rate}
\end{equation}
where $\psi_a$ and $\phi_a$ are the Euler angles for yaw and roll, $g$ is the acceleration due to gravity, and $\dot{s}$ is the speed of the aircraft.

The final stage of the ASG method uses the Euler angles and Euler angle rates along the segments to calculate the specific force and angular rates experienced by the aircraft body when following the smoothed path.
The ASG method in \cite{asg} shows that the true specific force is given by
\begin{equation}
    \boldsymbol{\nu}^{b} = T_n^b\begin{bmatrix} 
    \ddot{s}\cos(\psi_a) - \dot{s}\dot{\psi_a}\sin(\psi_a) \\ 
    \ddot{s}\sin(\psi_a) + \dot{s}\dot{\psi_a}\cos(\psi_a) \\
    -g \\
    \end{bmatrix} \label{eq:curvilinear_accel_final}
\end{equation}
and the angular rates are given by
\begin{equation}
    \boldsymbol{\omega}^b = \begin{bmatrix}
        \ddot{\psi_a}\frac{\dot{s}}{g}\cos^{2}\phi-\dot{\psi_a}\sin\theta_a \\
        \dot{\psi_a}\sin\phi \cos \theta_a \\
        \dot{\psi_a}\cos\phi \cos \theta_a
        \end{bmatrix}
\end{equation}
where $\ddot{s}$ is the acceleration in the direction of the path, and $T_n^b$ is the rotation matrix from the NED frame to the body frame.

The quantities generated from ASG are used as the nominal aircraft states, $\boldsymbol{\bar{x}_a}$, true specific force, $\boldsymbol{\nu^b}$, and true angular rates, $\boldsymbol{\omega^b}$.
These quantities are used by the INS model in Section \ref{sec:ins_model} to calculate the aircraft state covariance $C_{aa}$, which is used by the radar detection model from Section \ref{sec:radar_model} to calculate the variance of $P_D$, $\sigma_{pd}^2$.
The remainder of this paper shows how these elements are combined into a path planner that incorporates uncertainty in the aircraft and radar states.

\section{Error Budgets and Linear Covariance \label{sec:lincov}}
The INS model described in Section \ref{sec:ins_model} is used to estimate the uncertainty of a navigation state due to biased and noisy sensor measurements.
One approach to analyze the performance of the INS, is to generate error budgets that provide a graphical representation of the contribution of each source of uncertainty on the overall estimation uncertainty.
This section describes error budgets in more detail and derives a linear covariance model that is used to efficiently generate error budgets. 

\subsection{Error Budgets}
\paperplan{Error Budgets}
{
    \begin{itemize}
        \item Graphically show contribution of error (uncertainty) sources
        \item LinCov provides efficent generation of error budgets by avoiding Monte Carlo runs
        \item Define LinCov Models (truth and navigation)
        \item Describe approach to generating error budgets 
    \end{itemize}
}

\revisionC{Error budgets are generated from statistical information typically obtained through several Monte Carlo analyses \cite{maybeck_stochastic_1994,farrell_aided_2008}.
In this approach, a Monte Carlo analysis is first performed with all uncertainty sources activated.
Then the Monte Carlo analysis is repeated for each source of uncertainty where a single source is activated and all other sources are de-activated (i.e. noise samples set to zero). 
The ensemble statistics for each Monte Carlo analysis are used to estimate the time-varying navigation error covariance, $P_{true}$, due to the activated uncertainty source.}  

The error budget \revisionC{generation method} is used to analyze the effect of the sources of uncertainty on the overall navigation error covariance.
For the radar detection path planning application, the error budget of interest is the effect of the sources of uncertainty on the variability of $P_D$.
This introduces two additional uncertainty sources to be evaluated (i.e., radar position and radar constant).
The effect of each source of uncertainty on $P_D$ is analyzed by turning off all sources of uncertainty except the one being evaluated and calculating the variance of $P_D$ using \eqref{eq:pd_var}.
Once all sources of uncertainty have been evaluated, the variance of $P_D$ calculated during each test can be compared in an error budget. 

The $P_D$ error budget analysis includes eight Monte Carlo analyses with each requiring hundreds or thousands of navigation system simulations along the reference trajectory.
The high computational burden of this analysis prohibits its use in the proposed path planning framework.
An alternative method to compute the navigation error covariance is linear covariance analysis (LinCov) \cite{maybeck_stochastic_1994,christensen_2014,christensen2021closedloop}.
LinCov uses linear models for the truth and navigation states to efficiently generate the navigation error covariance along a reference trajectory.
This approach generates the same statistical information as a Monte Carlo analysis, but requires only a single simulation over the reference trajectory.
Thus, the error budget analysis described in this section can be performed with only eight simulations over the reference trajectory.
The following section describes the LinCov models used in this paper to obtain the true navigation error covariance $P_{true}$.

\subsection{Linear Covariance Model}
\paperplan{Linear Covariance Model}
{
    \begin{itemize}
        \item Linear modeling
        \item Open-loop navigation LinCov model
    \end{itemize}
}
This section describes the LinCov model associated with the truth and navigation models developed in Section \ref{sec:ins_model}.
The LinCov model forms an augmented state vector of the truth and navigation states and the associated linearized augmented system matrices.
The notation used in this section follows the development in \cite{christensen2021closedloop}.

The linearized truth state dispersion dynamics are determined by taking the Jacobian of the truth model defined in \eqref{eq:truth_model} to obtain
\begin{gather}
    \delta\dot{\boldsymbol{x}}=F_{x}\delta\boldsymbol{x}+B\boldsymbol{w}\label{eq:linmod4}
\end{gather}
where the uppercase letters represent the partial derivative of an equation taken with respect to the variable in the subscript. 
For example, $F_{\boldsymbol{x}}$ is the partial derivative of $\boldsymbol{f}(\cdot)$, as defined in \eqref{eq:truth_model}, with respect to the truth state, $\boldsymbol{x}$, evaluated at the nominal truth state, $\bar{\boldsymbol{x}}$. 
Given this definition, the matrices in \eqref{eq:linmod4} are defined as
\begin{equation}
    F_{x}=\begin{bmatrix}
    \boldsymbol{0} & \boldsymbol{I} & \boldsymbol{0} & \boldsymbol{0} & \boldsymbol{0}\\
    \boldsymbol{0} & \boldsymbol{0} & \left(\bar{T}_{b}^{n}\boldsymbol{\nu}^{b}\right)\times & \boldsymbol{0} & \boldsymbol{0}\\
    \boldsymbol{0} & \boldsymbol{0} & \boldsymbol{0} & \boldsymbol{0} & \boldsymbol{0}\\
    \boldsymbol{0} & \boldsymbol{0} & \boldsymbol{0} & -\frac{1}{\tau_{a}}\boldsymbol{I} & \boldsymbol{0}\\
    \boldsymbol{0} & \boldsymbol{0} & \boldsymbol{0} & \boldsymbol{0} & -\frac{1}{\tau_{g}}\boldsymbol{I}
    \end{bmatrix}
\end{equation}
and
\begin{equation}
    B=\left[\begin{array}{c}
    \boldsymbol{0}_{9\times6}\\
    I_{6\times6}
    \end{array}\right].
\end{equation}
The additive white noise vector is given by
\begin{equation}
    \boldsymbol{w}=\left[\begin{array}{c}
    \boldsymbol{w}_{a}\\
    \boldsymbol{w}_{g}
    \end{array}\right]
\end{equation}
where $\boldsymbol{w}_a$ and $\boldsymbol{w}_g$ are defined in \eqref{eq:we}-\eqref{eq:qg}.

The linearized navigation state dispersion dynamics are determined by taking the Jacobian of the navigation model defined in \eqref{eq:nav_model} to obtain
\begin{gather}
    \delta\dot{\hat{\boldsymbol{x}}}=\hat{F}_{\hat{x}}\delta\hat{\boldsymbol{x}}+\hat{F}_{\tilde{y}}C_{x}\delta\boldsymbol{x}+\hat{F}_{\tilde{y}}\boldsymbol{\eta}\label{eq:linmod5}
\end{gather}
where
\begin{equation}
    \hat{F}_{\hat{x}}=\begin{bmatrix}
    \boldsymbol{0} & \boldsymbol{I} & \boldsymbol{0} & \boldsymbol{0} & \boldsymbol{0}\\
    \boldsymbol{0} & \boldsymbol{0} & \left[\hat{\bar{T}}_{b}^{n}\left(\boldsymbol{\nu}^{b}-\hat{\bar{\boldsymbol{b}}}_{a}\right)\right]\times & -\hat{\bar{T}}_{b}^{n} & \boldsymbol{0}\\
    \boldsymbol{0} & \boldsymbol{0} & \boldsymbol{0} & \boldsymbol{0} & \hat{\bar{T}}_{b}^{n}\\
    \boldsymbol{0} & \boldsymbol{0} & \boldsymbol{0} & -\frac{1}{\tau_{a}}\boldsymbol{I} & \boldsymbol{0}\\
    \boldsymbol{0} & \boldsymbol{0} & \boldsymbol{0} & \boldsymbol{0} & -\frac{1}{\tau_{a}}\boldsymbol{I}
    \end{bmatrix} \label{eq:F_hat}
\end{equation}
\begin{equation}
    \hat{F}_{\tilde{y}}=\begin{bmatrix}
    \boldsymbol{0} & \boldsymbol{0}\\
    \hat{\bar{T}}_{b}^{n} & \boldsymbol{0}\\
    \boldsymbol{0} & -\hat{\bar{T}}_{b}^{n}\\
    \boldsymbol{0} & \boldsymbol{0}\\
    \boldsymbol{0} & \boldsymbol{0}
    \end{bmatrix}
\end{equation}
and
\begin{equation}
    C_{x}=\left[\begin{array}{cc}
    0_{6 \times 9} & I_{6\times6}\end{array}\right]_{6 \times 15}.
\end{equation}
The truth state update is linearized to obtain
\begin{equation}
    \delta\boldsymbol{x}_{k}^{+}=\delta\boldsymbol{x}_{k}^{-}. \label{eq:linmod7}
\end{equation}
The navigation state update equation defined in \eqref{eq:ekf_update} is also linearized to produce the dispersion update equation given by
\begin{equation}
    \delta\hat{\boldsymbol{x}}_{k}^{+}=\left(I_{16 \times 16}-\hat{K}_{k}\hat{H}_{\hat{x}}\right)\delta\hat{\boldsymbol{x}}_{k}^{-}+\hat{K}_{k}H_{x}\delta\boldsymbol{x}_{k}^{-}+\hat{K}_{k}\boldsymbol{\nu}_{k}\label{eq:linmod8}
\end{equation}
where the $H$ matrices depend on the measurement update type and are defined in \eqref{eq:ekf_p_meas}-\eqref{eq:ekf_psi_meas}.
Note that for the measurement models used in this paper, $H_{x}=\hat{H}_{\hat{x}}$.
    
The linearized truth and navigation models are combined to form the augmented system model.
The augmented system state vector for the linearized dispersion models is formed as
\begin{equation}
    \boldsymbol{X}=\left[\begin{array}{c}
    \delta\boldsymbol{x}\\
    \delta\hat{\boldsymbol{x}}
    \end{array}\right]
\end{equation}
and the augmented propagation and update equations are
\begin{equation}
    \dot{\boldsymbol{X}}=\mathcal{F}\boldsymbol{X}+\mathcal{G}\boldsymbol{\eta}+\mathcal{W}\boldsymbol{w}
\end{equation}
\begin{equation}
    \boldsymbol{X}_{k}^{+}=\mathcal{A}_{k}\boldsymbol{X}_{k}^{-}+\mathcal{B}_{k}\boldsymbol{\nu}_{k}
\end{equation}
    where
\begin{equation}
    \mathcal{F}=\left[\begin{array}{cc}
    F_{x} & 0\\
    \hat{F}_{\tilde{y}}C_{x} & \hat{F}_{\hat{x}}
    \end{array}\right]
\end{equation}
\begin{equation}
    \mathcal{G}=\left[\begin{array}{c}
    0\\
    \hat{F}_{\tilde{y}}
    \end{array}\right]
\end{equation}
\begin{equation}
    \mathcal{W}=\left[\begin{array}{c}
    B\\
    0_{16 \times 6}
    \end{array}\right]
\end{equation}
\begin{equation}
    \mathcal{A}_{k}=\left[\begin{array}{cc}
    I_{16 \times 16} & 0_{16 \times 16}\\
    K_{k}H_{x} & I_{16 \times 16}-K_{k}\hat{H}_{\hat{x}}
    \end{array}\right]
\end{equation}
\begin{equation}
    \mathcal{B}_{k}=\left[\begin{array}{c}
    0_{16 \times n_{z}}\\
    K_{k}
    \end{array}\right]
\end{equation}
and $n_z$ is dimension of the discrete-time measurement being processed.

Finally, the covariance propagation and update of the augmented system
is expressed as
\begin{align}
E\left[\dot{\boldsymbol{X}}\dot{\boldsymbol{X}}^{T}\right] = \dot{C}_{A} = \mathcal{F}C_{A}+C_{A}\mathcal{F}^{T}+\mathcal{G}S_{\eta}\mathcal{G}^{T}+\mathcal{W}S_{w}\mathcal{W}^{T}
\end{align}
\begin{align}
E\left[\boldsymbol{X}_{k}^{+}\boldsymbol{X}_{k}^{+T}\right] = C_{A}^{+} = \mathcal{A}_{k}C_{A}^{-}\mathcal{A}_{k}^{T}+\mathcal{B}_{k}R_{\nu}\mathcal{B}_{k}^{T}
\end{align}
where the power spectral density of the inertial measurements and process noise are
\begin{equation}
S_{\eta}=\left[\begin{array}{cc}
    Q_{\nu} & 0_{3\times3}\\
    0_{3\times3} & Q_{\omega}
    \end{array}\right]
\end{equation}
\begin{equation}
S_{w}=\left[\begin{array}{cc}
    q_{a}\boldsymbol{I} & 0_{3\times3}\\
    0_{3\times3} & q_{g}\boldsymbol{I}
    \end{array}\right]
\end{equation}
and $Q_{\nu}$, $Q_{\omega}$, $q_{a}$, and $q_{g}$ are defined in \eqref{eq:Q_PSD}-\eqref{eq:qg}.

The quantity of interest for this paper is the navigation estimation error. 
It is important to note that the navigation error covariance defined below is the true navigation error covariance, which may be different than the estimated navigation covariance from the Kalman filter. 
This quantity is extracted from the augmented covariance matrix via
\begin{equation}    
    P_{true}=\left[\begin{array}{ccc}
        -I & I_{16 \times 16} \end{array}\right]C_{A}\left[\begin{array}{ccc}
        -I & I_{16 \times 16} \end{array}\right]^{T}.
\end{equation}

Note that the error budget and linear covariance models developed in this section evaluate the variability in $P_D$ due to uncertainty in the aircraft INS and the radar states.
This approach evaluates knowledge errors due to the sources of uncertainty rather than vehicle dispersions that are affected by disturbances such as wind, or guidance and control designs.
\revision{This is convenient for the path planning application because it requires less computation than the vehicle dispersion analysis.
Furthermore, the navigation errors are a good approximation of the vehicle dispersions when the control authority of the vehicle is sufficient to follow the planned path in the presence of disturbances.}
\revision{If an analysis of the vehicle dispersions is desired, then a closed-loop LinCov model as developed in \cite{christensen2021closedloop} could be used in the framework presented in this paper.}
% \revision{Alternatively, the path planning framework presented in this paper can be used to evaluate the true state dispersions calculated by a closed-loop LinCov model as presented in \cite{christensen2021closedloop}.
% The true state dispersions incorporate the effect of the control system design and the disturbances acting on the aircraft.}

\section{Application: Visibility Graph Planner \label{sec:vgraph}}
\paperplan{VGraph Planning for Radar Detection}
{
    \begin{itemize}
        \item Polygon obstacles
        \item Graph of visible vertices with cost associated with length of segment
        \item Dijstras algorithm to find shortest path
        \item Variable RCS
        \begin{itemize}
            \item Need to have adjustment policy
            \item Describe adjustment policy
        \end{itemize}
    \end{itemize}
}

\paperplan{VGraph with Uncertainty}
{
    \begin{itemize}
        \item Use $\bar{P}_D + 3\sigma_{pd} < P_{D_{thresh}}$ as the objective
        \item Use similar adjustment policy but use CDF
    \end{itemize}
}

\paperplan{Planner Output}
{
    \begin{itemize}
        \item Aircraft states and covariances along planned path
        \item $\bar{P}_D$ and $\sigma_{pd}$ along path
    \end{itemize}
}
The methods presented in this paper can be used to inform a variety of radar detection path planning algorithms.
This section presents an application of these methods to a visibility graph path planner.
The following subsections describe the visibility graph path planner and an associated extension to incorporate the radar detection framework discussed in the previous sections.

\subsection{General Visibility Graph Path Planner}
\begin{figure}
\begin{centering}
\includegraphics[width=1\columnwidth]{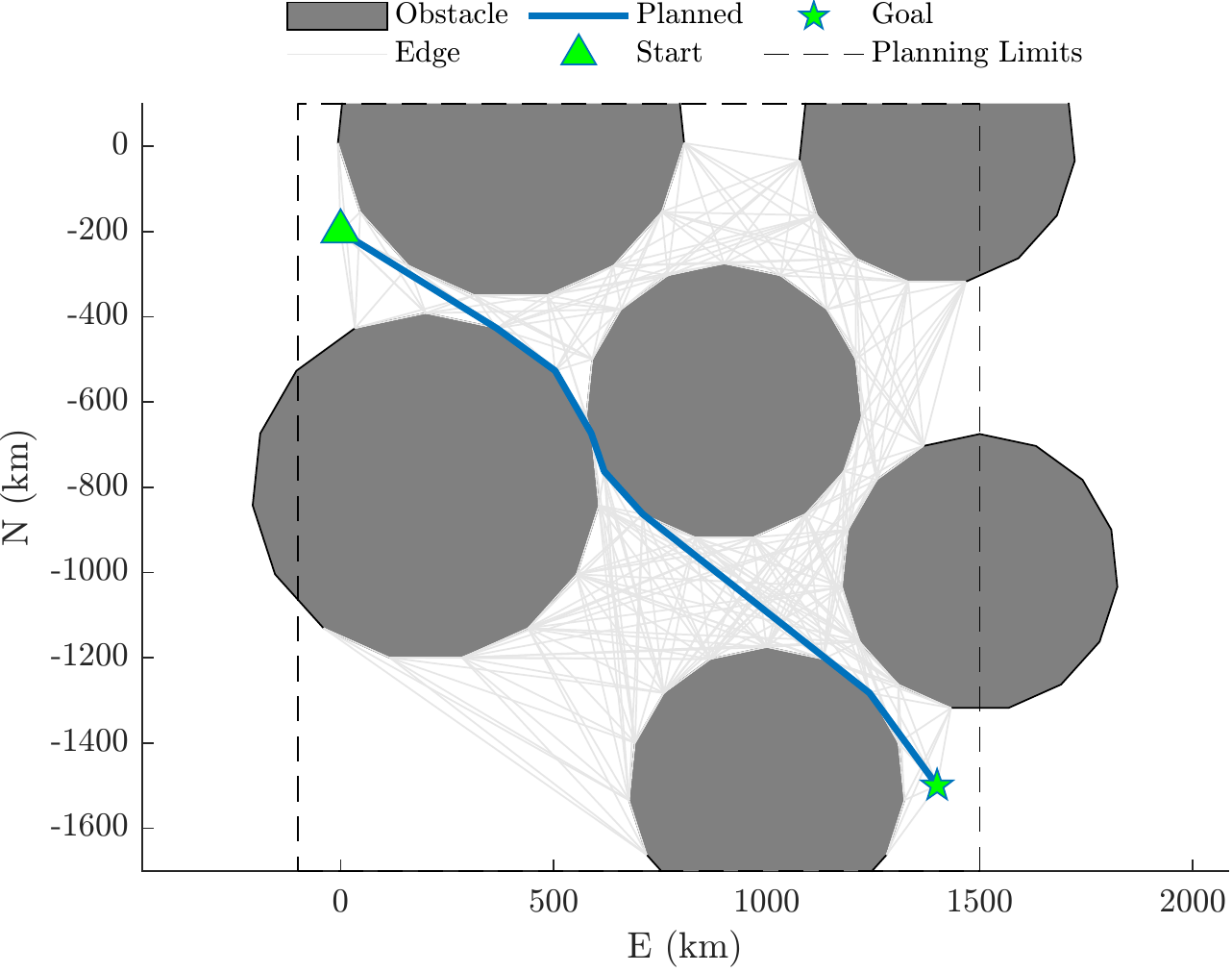}
\par\end{centering}
\caption{Visibility graph example with six polygon obstacles (dark gray) where the light gray lines represent the edges of the visibility graph\label{fig:Visibility-Graph-example}}
\end{figure}

The visibility graph path planner finds the shortest path between a start and goal location while navigating around obstacles.
\revision{The nodes in the graph are the start and goal points and the vertices of the obstacles.
Each node is connected by edges to every ``visible'' node.
In this context, a node is visible if a line segment to the node does not pass through an obstacle.
The edge cost is set to the Euclidean distance of the edge. 
The shortest path (the path with the least cost) between the goal and target vertices is found using a graph search algorithm such as Dijkstra's \cite{dijkstra_note_1959} or A* \cite{hart_astar}. 

Fig. \ref{fig:Visibility-Graph-example} shows an example of a visibility graph applied to a path planning problem where the gray lines represent the edges of the graph, and the gray polygons are the obstacles. 
The blue line is the shortest path between the start point (triangle) and the goal point (star) that was found using a shortest path algorithm.
The visibility graph method relies on the assumption that the shortest path through an obstacle field will touch the edges of the obstacles if a single straight segment between the start point and goal point is not valid.
}

The visibility graph planner described in this section is sufficient for radar detection planning with constant RCS and no uncertainty.
The following subsection describes an extension to the visibility graph path planning algorithm to incorporate the methods presented in this paper.
The resulting algorithm generates a path that maintains $P_D$ below a specified threshold and accounts for varying RCS and uncertainty in the aircraft and radar states.

\subsection{Visibility Graph Extension \label{sec:pdvg}}
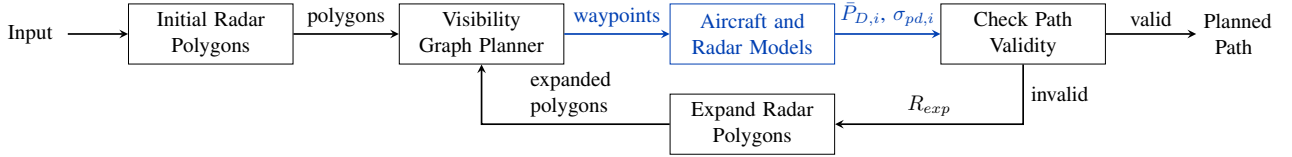
\begin{figure*}
    \begin{centering}
    \tikzstyle{startstop} = [rectangle, minimum width=2cm, minimum height=1cm,text centered, draw=black]
    \tikzstyle{process} = [rectangle, minimum width=2.5cm,  text width=2.5cm, minimum height=1cm, text centered, draw=black]
    \tikzstyle{arrow} = [thick,->,>=stealth]
    % \definecolor{bcolor}{rgb}{0.16, 0.32, 0.75}
    % \definecolor{bcolor}{rgb}{0.11, 0.22, 0.73}
    \definecolor{bcolor}{rgb}{0, 0.25, 0.70}
    \scalebox{0.8}{
    \begin{tikzpicture}[node distance=4.5cm]
        % Nodes
        \node (in) [text width=1cm, text centered] {Input};
        \node (start) [process, right of=in, xshift=-1.5cm] {Initial Radar Polygons};
        \node (pro1) [process, right of=start] {Visibility Graph Planner};
        \node (pro3) [process, color=bcolor, right of=pro1] {Aircraft and Radar Models};
        \node (pro4) [process, right of=pro3] {Check Path Validity};
        \node (pro5) [process, below of=pro3, yshift=3cm] {Expand Radar Polygons};
        \node (end) [text width=1cm, right of=pro4, text centered, xshift=-1cm] {Planned Path};
        % % Arrows
        \draw [arrow] (start) -- node[anchor=south] {polygons} (pro1);
        \draw [arrow, color=bcolor] (pro1) -- node[anchor=south] {waypoints}  (pro3);
        \draw [arrow, color=bcolor] (pro3) -- node[anchor=south] {$\bar{P}_{D,i}$, $\sigma_{pd,i}$} (pro4);
        \draw [arrow] (pro4) |- node[anchor=west, near start] {invalid} node[anchor=south, near end] {$R_{exp}$} (pro5);
        \draw [arrow] (pro5) -| node[anchor=south, near start, text width=1.5cm] {expanded polygons} (pro1);
        \draw [arrow] (pro4) -- node[anchor=south] {valid} (end);
        \draw [arrow] (in) -- (start);
    \end{tikzpicture}
    }
    \par\end{centering}
    \caption{Block diagram for the PDVG path planner. The Aircraft and Radar Models block in blue is expanded in Fig. \ref{fig:background_block_diagram}. \label{fig:planner_block_diagram}}
\end{figure*}

A block diagram for the $P_D$ visibility graph (PDVG) algorithm is shown in Fig. \ref{fig:planner_block_diagram} where the Aircraft and Radar Models block is expanded in Fig. \ref{fig:background_block_diagram}.
The algorithm represents the radar detection region with polygons so that a visibility graph problem can be formulated and solved to efficiently produce a candidate path.
The path is smoothed to provide continuous heading and curvature using the fillet method described in Section \ref{sec:ASG}. 
Once smoothed, the navigation covariance is propagated along the path using the INS model described in Section \ref{sec:ins_model}.
Finally, the radar detection model from Section \ref{sec:radar_model} is used to determine if the probability of detection stays within the user-specified requirements.
If the smoothed path satisfies the threshold, the algorithm is complete.
Otherwise, the radar polygons are expanded based on the radar detection information and a new plan is generated using expanded polygons.
The following paragraphs will describe these steps in more detail.

\subsubsection{Initial Radar Polygons}
The first step of the algorithm is to construct the initial radar detection polygons.
The detection polygons are parameterized by a position, number of vertices, and a radius. 
The position of the radar, $\boldsymbol{p_r^n}$, is the center of the detection polygon. 
The number of vertices is a design decision by the user and the radius is determined by solving \eqref{eq:pd_approx} and \eqref{eq:SNR} for $R$ as
\begin{equation}
	R = \left(\frac{c_r \sigma_{r}}{k \left(\textrm{erfcinv}\left(2P_{D,\textrm{init}}\right)-\sqrt{-\log{P_{fa}}}\right)^2 - 0.5}\right)^\frac{1}{4}. \label{eq:range_pd}
\end{equation}
where $P_{D,\textrm{init}}$ is a $P_D$ value used to construct the polygons, and $\sigma_r$ is a nominal RCS value for the aircraft.
Typically, $P_{D,\textrm{init}}$ is the same or slightly lower than the mission $P_D$ threshold $P_{DT}$, and $\sigma_r$ is the average RCS value of the chosen RCS model.

\subsubsection{Visibility Graph Planner}
The visibility graph planner uses the radar polygons to build a visibility graph and calculates the shortest path from the start point to the goal point.
The resulting path is a series of waypoints that mark a path through the planning region.

\subsubsection{Aircraft and Radar Models}
The Aircraft and Radar Models are used to generate samples of $P_D$ and $\sigma_{pd}$ from a series of waypoints.
The ASG method from \cite{asg} and summarized in Section \ref{sec:ASG} generates a flyable trajectory from the waypoints.
The flyable trajectory includes nominal aircraft states, $\boldsymbol{\bar{x}_a}$, and nominal IMU measurement samples ($\boldsymbol{\nu^b}$, $\boldsymbol{\omega^b}$) along the path.
The inertial navigation model uses the nominal aircraft states and IMU measurements to compute the aircraft state covariance, $C_{aa}$, along the candidate path using the methods described in Section \ref{sec:ins_model}.

The nominal aircraft state and aircraft state covariance are used to calculate the time-varying nominal $\bar{P}_D(t)$ using \eqref{eq:pd_approx} and the variance of $P_D$, $\sigma_{pd}^2(t)$, using \eqref{eq:pd_var}.
The variance of $P_D$ along the nominal trajectory, as defined in \eqref{eq:pd_var}, is modified for the multiple-radar scenario such that, for the $i^{th}$ radar, $\sigma_{pd}$ is given by
\begin{align}
    \sigma^2_{pd,i}(t) = & A_{Pa,i}(t)C_{aa}(t)A_{Pa,i}(t)^\intercal \nonumber \\
                    & \qquad + A_{Pr,i}(t)C_{rr,i}A_{Pr,i}(t)^\intercal. \label{eq:pd_var_t}
\end{align}
% where the Jacobians $A_{Pa}$ and $A_{Pr}$ are provided in Appendix \ref{app:rad_jacs}, the aircraft state covariance $C_{aa}$ is calculated using the INS model described in Section \ref{sec:ins_model}, and the radar state covariance $C_{rr}$ is considered constant in this paper.

% Let $\boldsymbol{\bar{x}_a}(t_k)$ represent the nominal aircraft state at time $k$, $t_k$.
% Then the $\bar{P}_D^i(t_k)$ is the instantaneous probability of detection of the aircraft by the $i^{th}$ radar at time $t_k$ using \eqref{eq:pd_approx} and \eqref{eq:SNR_exp}. 
% Then let $C_{aa}(t_k)$ represent the time-varying aircraft state covariance at $t_k$ and $C^i_{rr}$ represent the radar state covariance for the $i^{th}$ radar. 
% Then $A_{Pa}^{i}(t_k)$ represents the Jacobian of $P_{D}$ \eqref{eq:pd_approx} with respect to the aircraft state $\boldsymbol{x_a}$ for the $i^{th}$ radar at time $k$.
% Similarly, $A_{Pr}^{i}(k)$ represents the Jacobian of $P_{D}$ \eqref{eq:pd_approx} with respect to the radar state, $\boldsymbol{x_r}$, for the $i^{th}$ radar at time $k$.
% Then the variance of $P_{D}$ at time $k$ for the $i^{th}$ radar is given by
% \begin{align}
% \sigma_{pd}^{i}(k)^{2} = & A_{Pa}^{i}(k)C_{aa}(k)A_{Pa}^{i}(k)^\intercal \nonumber \\
%                 & \qquad + A_{Pr}^i(k)C^i_{rr}A^i_{Pr}(k)^\intercal. \label{eq:pd_variance}
% \end{align}

\subsubsection{Check Path Validity}
The $\bar{P}_{D,i}$ and $\sigma_{pd,i}$ are used to determine if the candidate path is valid.
For this application, a path is considered valid if the nominal $P_D$ plus a multiple of $\sigma_{pd}$ stays below a specified $P_D$ threshold, $P_{DT}$, for all time as described below.
Let $m_{\sigma}$ be a mission planner specified multiple of the standard deviation of $P_{D}$. 
Then a path is considered valid if 
\begin{eqnarray}
\bar{P}_{D,i}(t)+m_{\sigma}\sigma_{pd,i}(t) & < & P_{DT} \qquad\forall\;t,i. \label{eq:valid_check}
\end{eqnarray}
Note that the $\sigma_{pd}$ value calculated with the linearized radar detection model using \eqref{eq:pd_var} represents the standard deviation of a Gaussian distribution. 
So the $m_\sigma$ multiple follows the empirical rule such that one, two, and three deviations will contain 68\%, 95\%, and 99.7\% of samples from the distribution, respectively.

\subsubsection{Radar Polygon Expansion}
If a candidate path passes the validity check in \eqref{eq:valid_check} for all time and every radar, the algorithm terminates and the candidate path is provided as the planned path.
Otherwise, the detection statistics are used to expand the radar polygons. 

\begin{figure}
    \begin{centering}
        \input{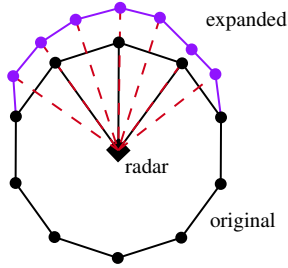}
        \par\end{centering}
        \caption{Radar expansion diagram. The black polygon represents the original polygon, the blue segments represent the polygon expansion, and the dashed red lines show the expansion vertices. Vertices of the original and expanded polygons are represented by black and blue circles, respectively. \label{fig:radar_expan}}
\end{figure}

The radar polygon expansion component expands the polygons in areas where the candidate path was invalid according to \eqref{eq:valid_check}.
The polygon expansion is shown graphically in Fig. \ref{fig:radar_expan} and is accomplished in two steps as described in the following paragraphs.

First, an expansion range is determined using $\bar{P}_{D,i}$ and $\sigma_{pd,i}$ for all samples where the path is invalid according to \eqref{eq:valid_check}.
The objective of the expansion range is to move the nominal path away from the radar to pass the validity check in \eqref{eq:valid_check}.
This is accomplished by determining an expansion $P_D$, $P_{D, exp}$, that when added to $m_{\sigma}\sigma_{pd}$ stays below $P_{DT}$. 
Let $P_{D,exp}$ for the $i^{th}$ radar be defined as
\begin{eqnarray}
P_{D,exp,i}(t) & = & \textrm{max}\left(P_{DT}-m_{\sigma}\sigma_{pd,i}(t)\textrm{, } 1e^{-3}\right)\label{eq:p_adj}
\end{eqnarray}
where the $\textrm{max}()$ function is used to ensure that $P_{D,exp,i}$ is positive and avoids the asymptote of the $\textrm{erfcinv}()$ function at zero.
Then the associated expansion range $R_{exp,i}(t)$ is determined by solving \eqref{eq:range_pd} with $P_{D,\textrm{init}} = P_{D,exp,i}(t)$ and $\sigma_r = \sigma_{r,i}(t)$. 

\revisionC{Second, the expansion ranges calculated in the first step are applied to the closest radar polygon vertices.
Then vertices are added to the polygon near the expanded vertices with the same expansion range as represented in Fig. \ref{fig:radar_expan}.}

The expanded radar polygons are provided to the visibility graph component and the algorithm continues until a valid path is found.
The iterative nature of this algorithm allows the visibility graph component to provide a new candidate path according to the expanded obstacles.

\section{Results \label{sec:Results}}
\paperplan{Results}
{
    \begin{itemize}
		\item Discuss path planning environment
		\item Introduce Scenarios
		\item Comment on tables of common parameters and RCS models
    \end{itemize}
}

\begin{figure}
	\centering
    \begin{subfigure}{0.49\columnwidth}
		\centering
		\includegraphics[width=\textwidth]{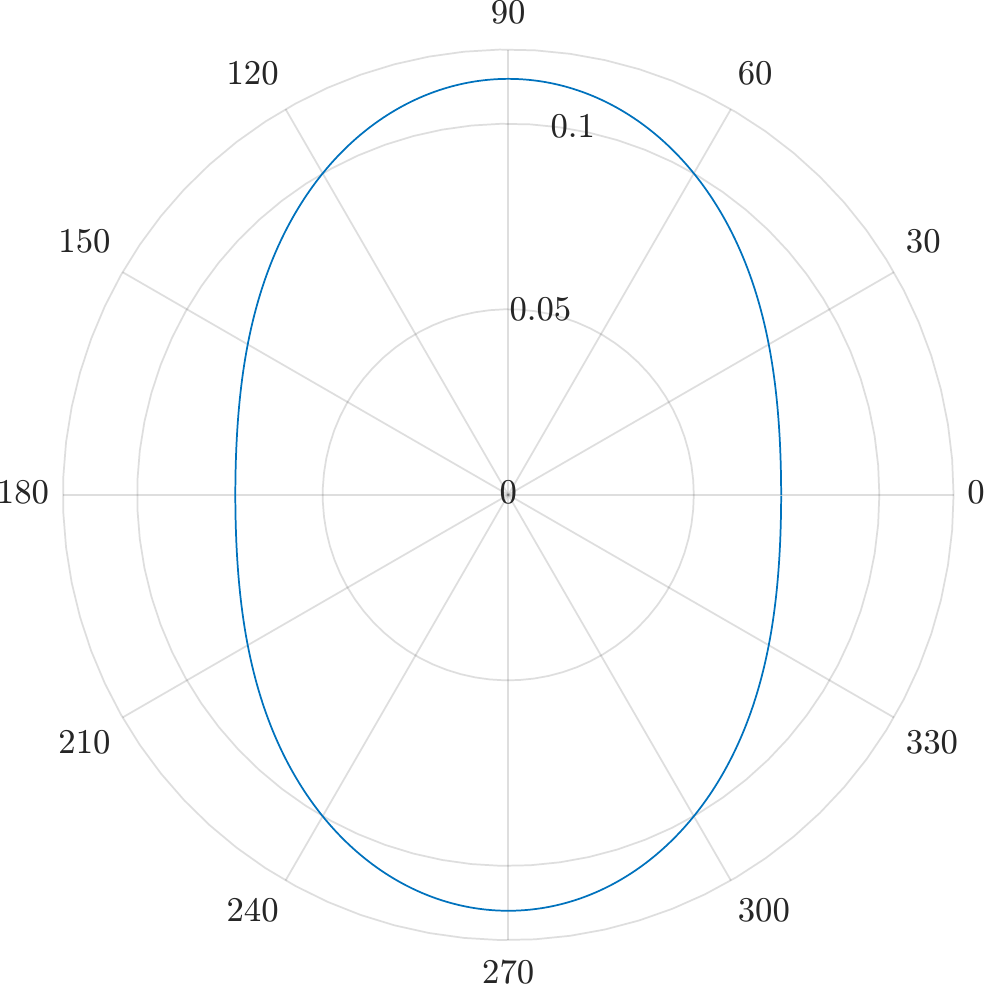}
		\caption{RCS Azimuth ($\alpha$)}
	\end{subfigure}
	\begin{subfigure}{0.49\columnwidth}
		\centering	
		\includegraphics[width=\textwidth]{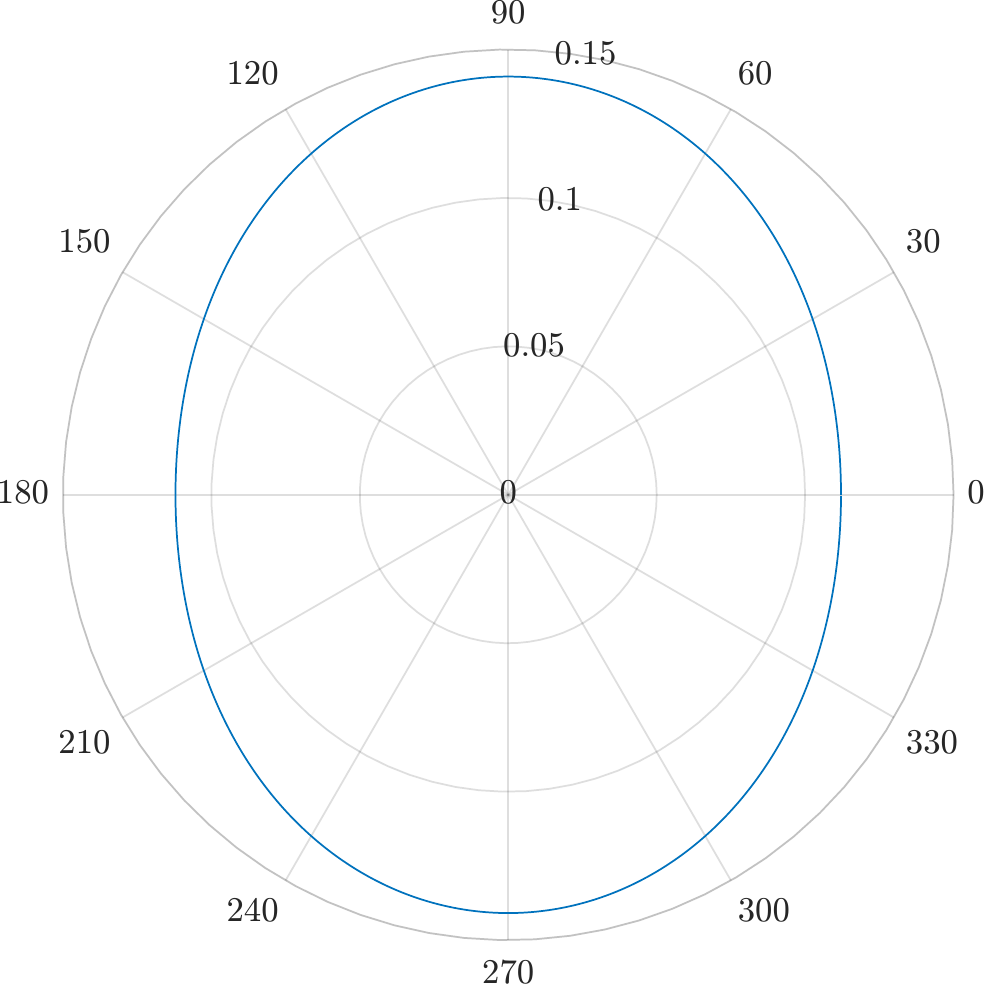}
		\caption{RCS Elevation ($\phi$)}		
	\end{subfigure}
	\caption{Ellipsoid RCS model as a function of RCS azimuth and elevation angles.\label{fig:rcs_model}}
\end{figure}

The results in this section illustrate the performance of the PDVG path planner described in Section \ref{sec:pdvg} for three scenarios.
The planner seeks to satisfy \eqref{eq:valid_check} with a threshold of $P_{DT}=0.1$ and $m_\sigma=3$.
The results for each scenario include a 2D map of the planning region, $P_D$ calculated along the planned trajectory, and an error budget for the sources of uncertainty.
The common parameters for the three scenarios are provided in Table \ref{tab:RadarParamsResults} and the ellipsoid RCS model as a function of azimuth and elevation angles are shown in Fig. \ref{fig:rcs_model}.

\begin{table}
    \begin{centering}
    \caption{Common parameters for the scenarios in this section. \label{tab:RadarParamsResults}}
    \par\end{centering}
    \centering{}%
    {\renewcommand{\arraystretch}{1.25}
    \begin{tabular}{c|c|l}
    \normalsize{\textbf{Param}} & \normalsize{\textbf{Value}} & \normalsize{\textbf{Description}}\tabularnewline
    \hline 
    $a$ & 0.18 $\textrm{m}$ & Ellipsoid RCS forward axis length \tabularnewline
    $b$ & 0.17 $\textrm{m}$ & Ellipsoid RCS side axis length\tabularnewline
    $c$ & 0.20 $\textrm{m}$ & Ellipsoid RCS up axis length \tabularnewline
    ${\sigma_{pr}}$ & 500/3 $\textrm{m}$ & Radar position std. dev. \tabularnewline
    $\sigma_{cr}$ & 2/3 J$\textrm{m}^2/^{\circ}$K & Radar constant std. dev. \tabularnewline
    $p_{ad}$ & -3.5 $\textrm{km}$ & Nominal aircraft position - "D" axis \tabularnewline
    $\boldsymbol{p_{r,1}^n}$ & $\boldsymbol{0_{3\times1}}$ $\textrm{km}$ & Radar 1 nominal position (NED) \tabularnewline
    $\boldsymbol{p_{r,2}^n}$ & [-650, 900, 0]$^\intercal$ $\textrm{km}$ & Radar 2 nominal position (NED) \tabularnewline
	$c_r$ & $164.7$ J$\textrm{m}^2/^{\circ}$K  & Nominal radar constant \tabularnewline
    $P_{fa}$ & $1\times 10^{-9}$ & Probability of false alarm \tabularnewline
	$\boldsymbol{p^n_{\textrm{start}}}$ & [-100, -700, -3.5]$^\intercal$ km & Start position \tabularnewline 
	$\boldsymbol{p^n_{\textrm{goal}}}$ & [-400, 1650, -3.5]$^\intercal$ km & Goal position \tabularnewline
	$P_{D,\textrm{init}}$ & 0.1 & $P_D$ used for initial vgraph \tabularnewline
	$\sigma_{r,\textrm{init}}$ & 0.09 $\textrm{m}^2$ & RCS used for initial vgraph \tabularnewline
	$n_p$ & 30 & \# of radar polygon vertices \tabularnewline 
	$P_{DT}$ & 0.1 & $P_D$ threshold for planner \tabularnewline
	$\sigma_n$ & 1/3 m & North position noise std. dev. \tabularnewline
	$\sigma_e$ & 1/3 m & East position noise std. dev. \tabularnewline
	$\sigma_d$ & 1 m & Down position noise std. dev. \tabularnewline
	$\sigma_h$ & 0.1/3 m & Altitude noise std. dev. \tabularnewline
	$\sigma_{\psi}$ & 0.1/3 deg. & Heading noise std. dev. \tabularnewline
	\end{tabular}
    }
\end{table}

The three scenarios include a start point, goal point, radar systems and ``planning limits'' in which the planned path must remain.
Fig. \ref{fig:res_scenario1} shows the map for Scenario 1 with the start (triangle) and goal (star) points.
The diamonds represent the position of the radar systems and the gray circles surrounding the diamonds show the radar detection regions.
The radius of the radar detection regions are determined using \eqref{eq:range_pd} with $\sigma_r = 0.15 \; \textrm{m}^2$ and $P_{D,\textrm{init}} = 0.01$.
These quantities were chosen so that the radar detection region would reflect a worst-case scenario for the RCS and a very low $P_D$ value.
The scenarios in this section will compare the effect of the IMU grade (i.e. industrial, tactical) on the uncertainty in $P_D$ and illustrate how this influences the path planner.
The IMU grade parameters used in this section are provided in Table \ref{tab:imu_grade}.

\begin{table}[]
	\centering
	\caption{Parameters for tactical and navigation grade IMU's. \label{tab:imu_grade}}
	\begin{tabular}{c|c|c|c|c}
	 \textbf{IMU} & $\boldsymbol{3\sqrt{q_a}}$ & $\boldsymbol{3\sigma_{a,ss}}$ & $\boldsymbol{3\sqrt{q_g}}$ & $\boldsymbol{3\sigma_{g,ss}}$ \\ 
	 \textbf{Grade} & \textbf{(m/s/}$\boldsymbol{\sqrt{\textrm{hr}}}$\textbf{)} & \textbf{(}$\boldsymbol{g}$\textbf{)} & \textbf{(deg./}$\boldsymbol{\sqrt{\textrm{hr}}}$\textbf{)} & \textbf{(deg./hr.)} \\
		\hline
		Industrial & 0.1 & 0.001 & 0.2 & 10 \\
		Tactical &  0.03 & 0.0001 & 0.05 & 1 
	\end{tabular}
\end{table}

\subsection{Scenario 1}
\paperplan{Scenario 1}
{
    \begin{itemize}
		\item Ellipsoid RCS, tactical grade IMU, no GPS-denied region
		\item 2D Plot showing path through radar detection regions
        \item PD Plot with $\bar{P}_D + 3\sigma_{pd}$
        \item Error budget bar plot
    \end{itemize}
}

\begin{figure}
	\begin{centering}
		\includegraphics[width=1\columnwidth]{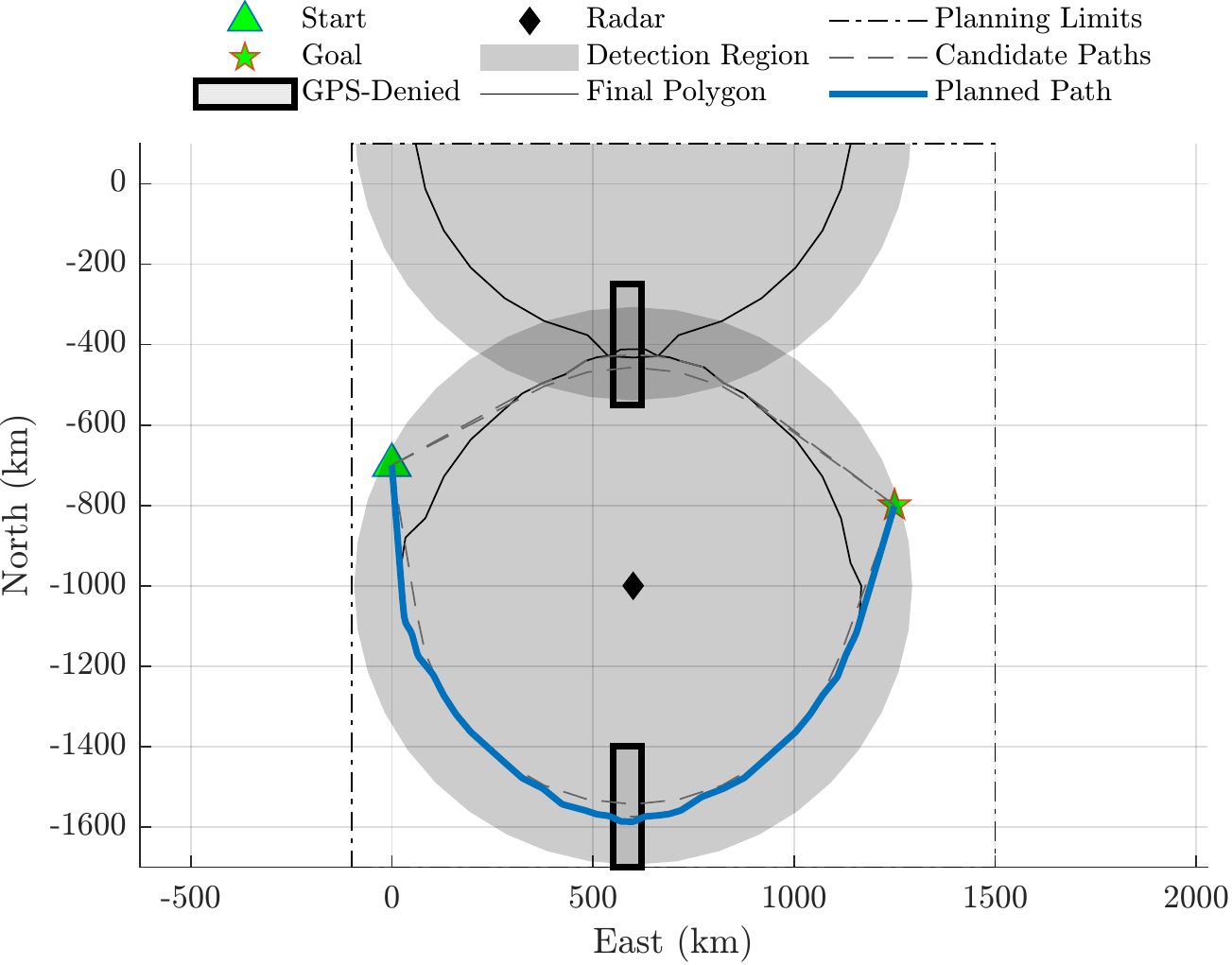}
		\par\end{centering}
	\caption{Visibility graph path planner results for Scenario 1 with two radar systems (diamonds), radar polygons (black lines), GPS-Denied regions (black rectangles), candidate paths (dashed gray lines), and the final planned path (blue line).\label{fig:res_scenario1}}
\end{figure}

The first scenario shows a planned path for an aircraft equipped with a industrial-grade IMU.
Fig. \ref{fig:res_scenario1} shows a map with the planned and candidate paths of the path planner.
The black polygons show the final state of the radar polygons used in the visibility graph planner.
The initial candidate path shows that the shortest path is between the two radar units.
As the aircraft travels through a GPS-Denied region, position and heading measurements are made unavailable to the INS.
This causes the aircraft pose covariance to grow which contributes to an increase in $\sigma_{pd}$.
The radar polygons are adjusted in areas where \eqref{eq:valid_check} is \revisionC{not satisfied} according to the adjustment policy described in Section \ref{sec:vgraph}.
The adjusted polygons overlap making the path between the radar units infeasible.
The final planned path goes around the bottom of the lower radar unit but remains within the gray radar detection region. 

\begin{figure}
	\begin{centering}
		\includegraphics[width=1\columnwidth]{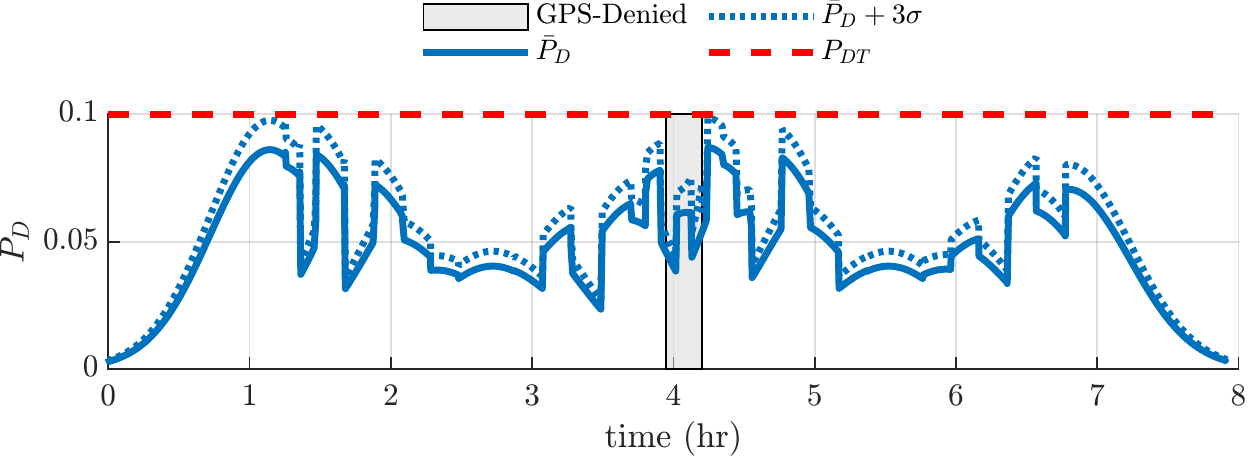}
		\par\end{centering}
	\caption{$P_D$ results for Scenario 1 with the $P_D$ threshold $P_{DT}=0.1$, nominal $\bar{P}_D$, and $\bar{P}_D + 3\sigma_{pd}$. The planner successfully finds a path through the radar detection region that maintains $\bar{P}_D + 3\sigma_{pd} < P_{DT}$. \label{fig:pd_res_scenario1}}
\end{figure}

Fig. \ref{fig:pd_res_scenario1} shows the $P_D$ results for Scenario 1.
This indicates that $\bar{P}_D + 3\sigma_{pd}$ stays below the $P_D$ threshold of 0.1 as required by \eqref{eq:valid_check}.
Note that the areas of highest detection risk and largest $\sigma_{pd}$ occur when the radar is detecting the side of the aircraft (i.e. $\alpha\approx90$, $270$ deg.).
This is expected as these azimuth angles are associated with the largest RCS values and highest degree of variability (see Fig. \ref{fig:rcs_model}).
Also note that at its peak ($t \approx 4.2$ hr.), $3\sigma_{pd} \approx 0.018$ which is $34.4\%$ of the nominal $\bar{P}_D$ at that time.
\revisionC{These results indicate that there is substantial variability in $P_D$ due to uncertainty in the aircraft pose, radar position, and radar parameters.}

\begin{figure}
	\begin{centering}
		\includegraphics[width=1\columnwidth]{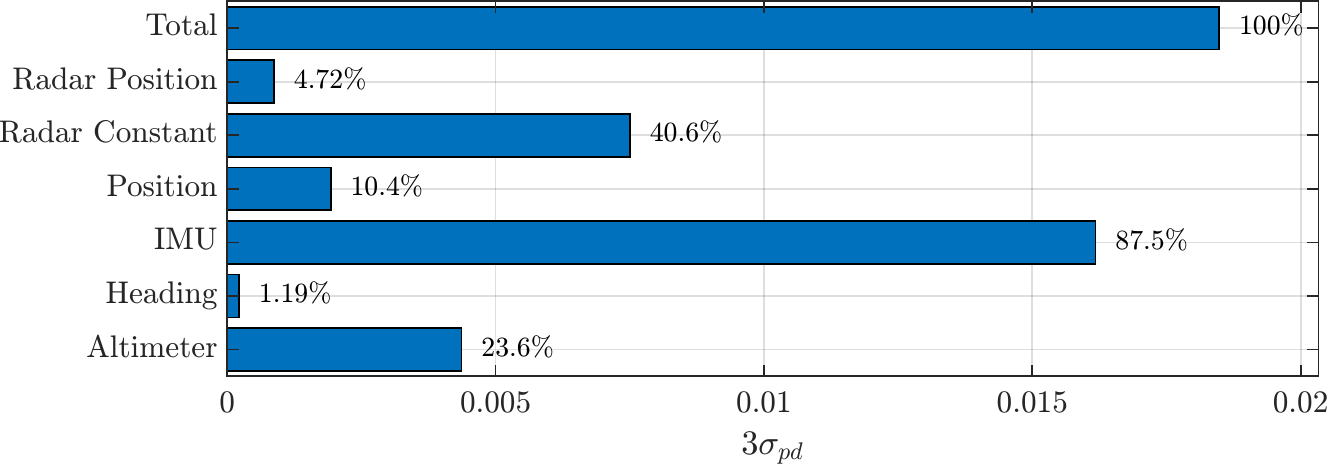}
		\par\end{centering}
	\caption{Error budget results for Scenario 1 at $t=4.2$ hours. This shows the $3\sigma_{pd}$ value at the stated time when each source of uncertainty is isolated. The Total line shows the $3\sigma_{pd}$ value of the LinCov model with all the sources of uncertainty activated. The percentages to the side of each bar indicate the percent of the Total uncertainty. \label{fig:eb_scenario1}}
\end{figure}

Fig. \ref{fig:eb_scenario1} shows an error budget for the sources of uncertainty at $t=4.2$ hours using the methods presented in Section \ref{sec:lincov}.
This time was chosen because it is when the aircraft is about to exit the GPS-Denied region and the aircraft pose uncertainty is the largest.
The error budget indicates that the uncertainty in the IMU is the primary driver of uncertainty in the variability of $P_D$.
This indicates that to reduce the uncertainty in $P_D$ the uncertainty in the IMU measurements must be reduced which will be examined in Scenario 2.

\begin{figure}
	\begin{centering}
		\togglefig{
		\includegraphics[width=1\columnwidth]{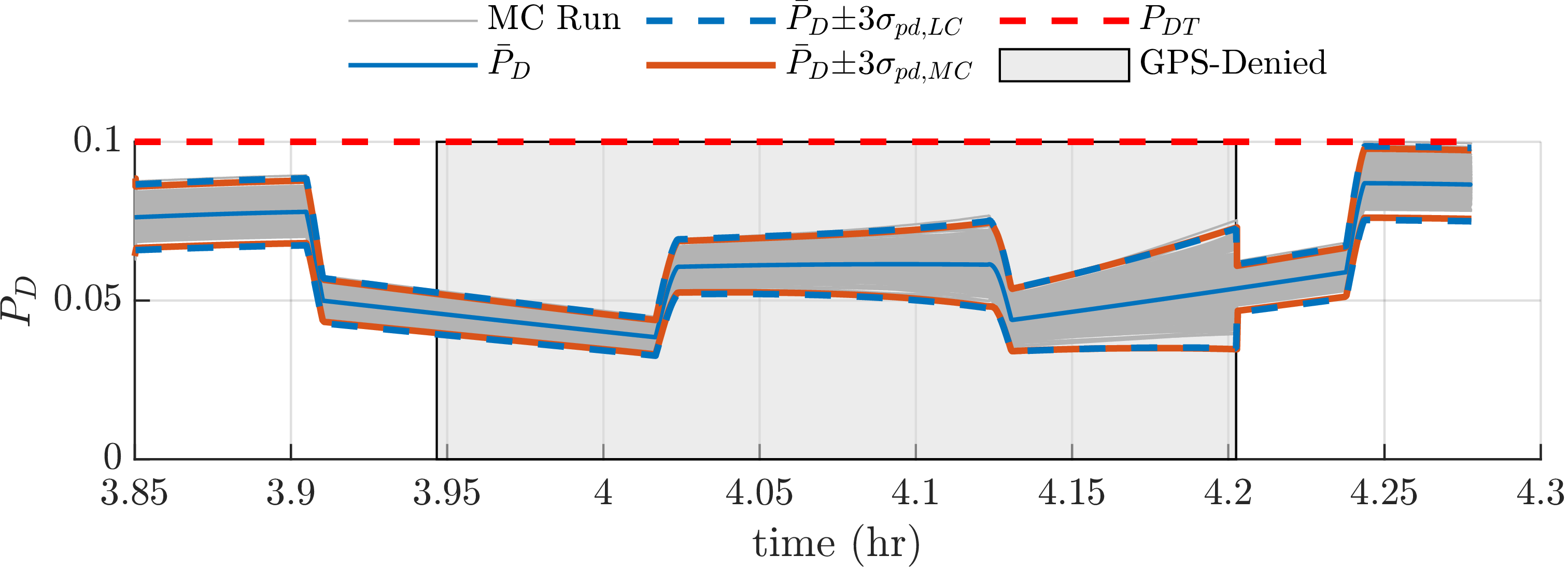}
		\includegraphics[width=1\columnwidth]{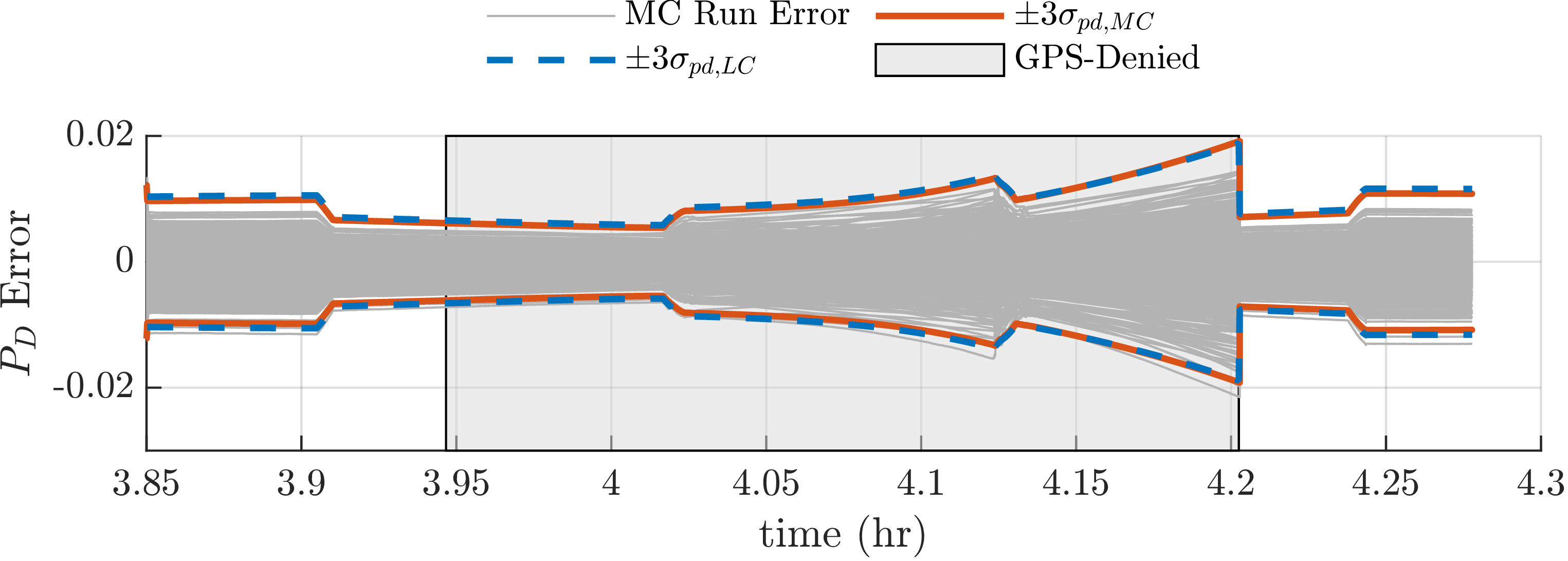}
		}
		\par\end{centering}
	\caption{$P_D$ Monte Carlo results with 500 runs for the path segments around the first radar. The gray lines indicate $P_D$ results for each Monte Carlo run with the $P_D$ error results in the bottom figure. The $P_D$ error is determined by subtracting the Monte Carlo run $P_D$ from $\bar{P}_D$. \label{fig:pd_mc_res_scenario1}}
\end{figure}

The LinCov analysis that is used to generate the error budget in Fig. \ref{fig:eb_scenario1} is expected to provide the same statistical information as a Monte Carlo analysis.
Fig. \ref{fig:pd_mc_res_scenario1} shows Monte Carlo results for a subset of the planned trajectory near the GPS-Denied region below the lower radar unit.
The gray lines in Fig. \ref{fig:pd_mc_res_scenario1} represent the $P_D$ results for each of the 500 Monte Carlo runs.
\revisionC{As expected, the} gray lines mostly stay within the $3\sigma_{pd}$ lines and the plots show agreement between the LinCov 3$\sigma$ and the Monte Carlo 3$\sigma$.
\revisionC{Agreement between the Monte Carlo and LinCov results serve to validate the linear approximations made in the LinCov framework.}
There are minor deviations throughout the trajectory due to linearization errors (i.e. minor bias between $t=4.15$-$4.25$ hrs.), but the deviations are negligible for the path planning scenario presented. 

\subsection{Scenario 2}
\begin{figure}
	\begin{centering}
		\includegraphics[width=1\columnwidth]{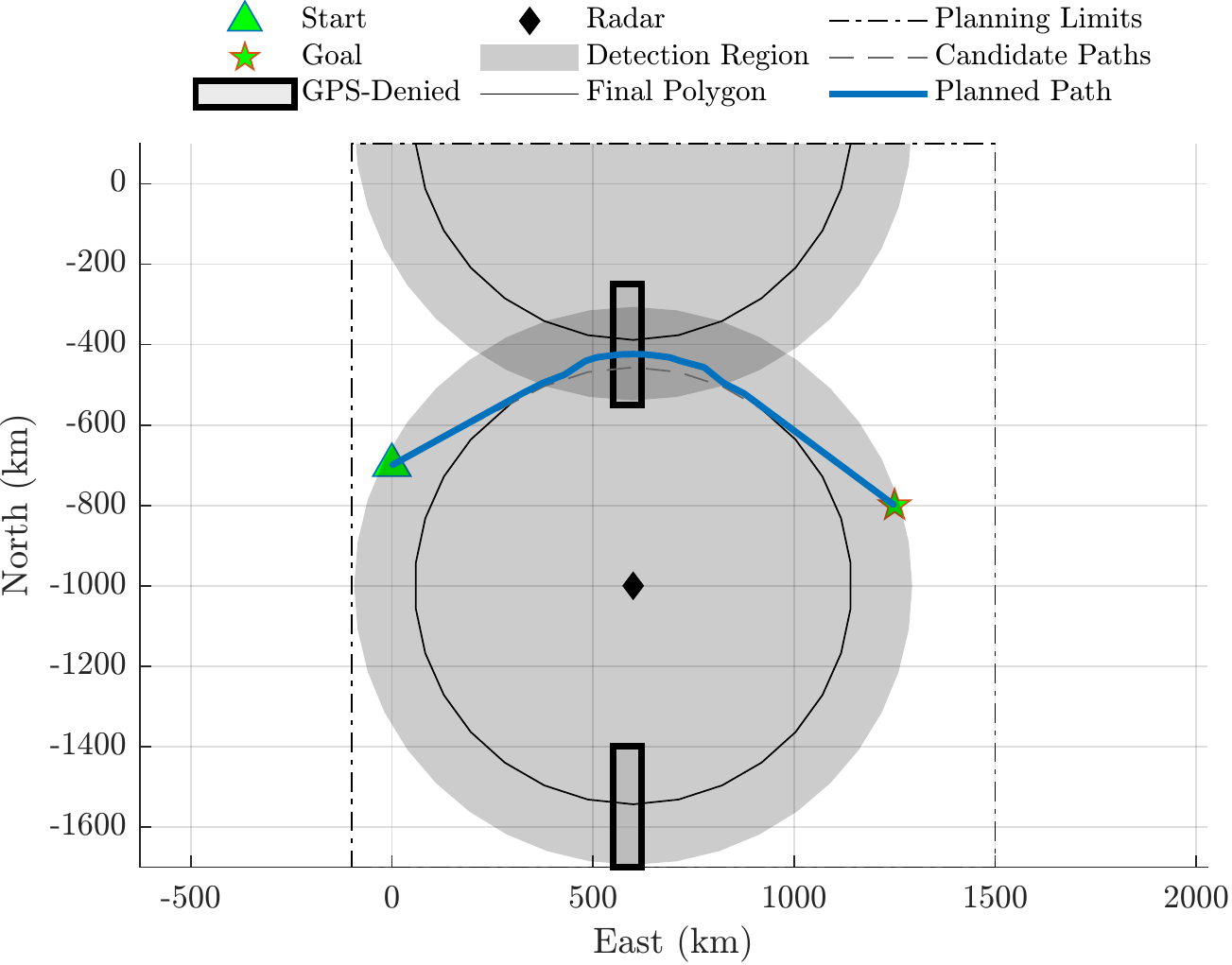}
		\par\end{centering}
	\caption{Visibility graph path planner results for Scenario 2 with two radar systems (diamonds), radar polygons (black lines), GPS-Denied regions (thick black rectangles), candidate paths (dashed gray lines), and the final planned path (blue line).\label{fig:res_scenario2}}
\end{figure}

\paperplan{Scenario 2}
{
    \begin{itemize}
		\item GPS-denied regions
		\item Ellipsoid RCS, tactical grade IMU
		\item No feasible path between radar
		\item 2D Plot showing path through radar detection regions
        \item PD Plot with $\bar{P}_D + 3\sigma_{pd}$
        \item Error budget combined with Scenario 3
    \end{itemize}
}
The second scenario uses the same radar and aircraft configuration as Scenario 1, except the industrial grade IMU is replaced by a tactical grade IMU (see Table \ref{tab:imu_grade}).
Fig. \ref{fig:res_scenario2} shows the 2D map for Scenario 2 where the final planned path goes between the radar units.
In contrast to Scenario 1 with the industrial grade IMU, the tactical grade IMU makes the shorter path between the radar feasible by reducing the aircraft pose covariance.

\begin{figure}
	\begin{centering}
		\includegraphics[width=1\columnwidth]{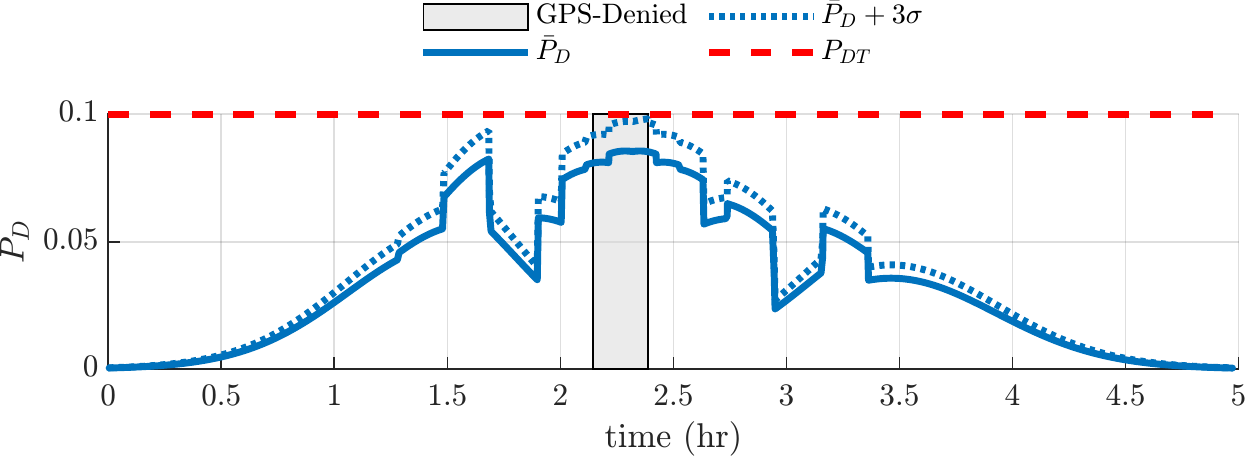}
		\par\end{centering}
	\caption{$P_D$ results for Scenario 2 with the $P_D$ threshold $P_{DT}=0.1$, nominal $\bar{P}_D$, and $\bar{P}_D + 3\sigma_{pd}$. The planner successfully finds a path through the radar detection region that maintains $\bar{P}_D + 3\sigma_{pd} < P_{DT}$. The effect of the GPS-Denied region is the increase in $3\sigma_{pd}$ near $t=2.4$ hours. \label{fig:pd_res_scenario2}}
\end{figure}

Fig. \ref{fig:pd_res_scenario2} shows the $P_D$ results for Scenario 2.
The graph shows that the PDVG planner maintains $\bar{P}_D + 3\sigma_{pd}$ below the $P_{DT}$ threshold of 0.1.
Note that at its peak ($t \approx 2.38$ hr.), $3\sigma_{pd} \approx 0.013$ which is $15.2\%$ of the nominal $\bar{P}_D$ at that time.
\revisionC{This is a significant portion of the nominal $P_D$ that must be considered for radar detection path planning in the presence of uncertainty, but is smaller than the variation calculated in Scenario 1.} 

\begin{figure}
	\begin{centering}
		\includegraphics[width=1\columnwidth]{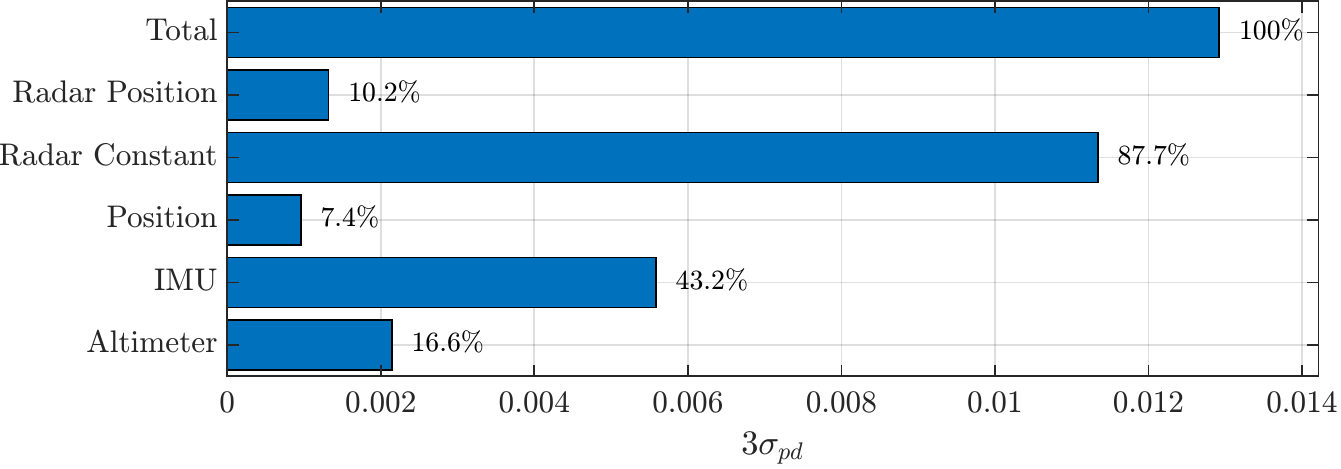}
		\par\end{centering}
	\caption{Error budget results for Scenario 2 at $t=2.38$ hours. The Total line shows the $3\sigma_{pd}$ value of the LinCov model with all the sources of uncertainty activated. The percentages to the side of each bar indicate the percent of the Total uncertainty for each of the sources of uncertainty. \label{fig:eb_scenario2}}
\end{figure}

The error budget for Scenario 2 at $t=2.38$ hours, which is just before the aircraft exits the GPS-Denied region, is provided in Fig. \ref{fig:eb_scenario2}.
The graph indicates that the radar constant is the primary driver of uncertainty in $P_D$ followed by the IMU.
Contrast these results with the error budget in Scenario 1 (Fig. \ref{fig:eb_scenario1}) to observe the reduction of the variability of $P_D$ due to the IMU grade improvement.

\subsection{Scenario 3}
\begin{figure}
	\begin{centering}
		\includegraphics[width=1\columnwidth]{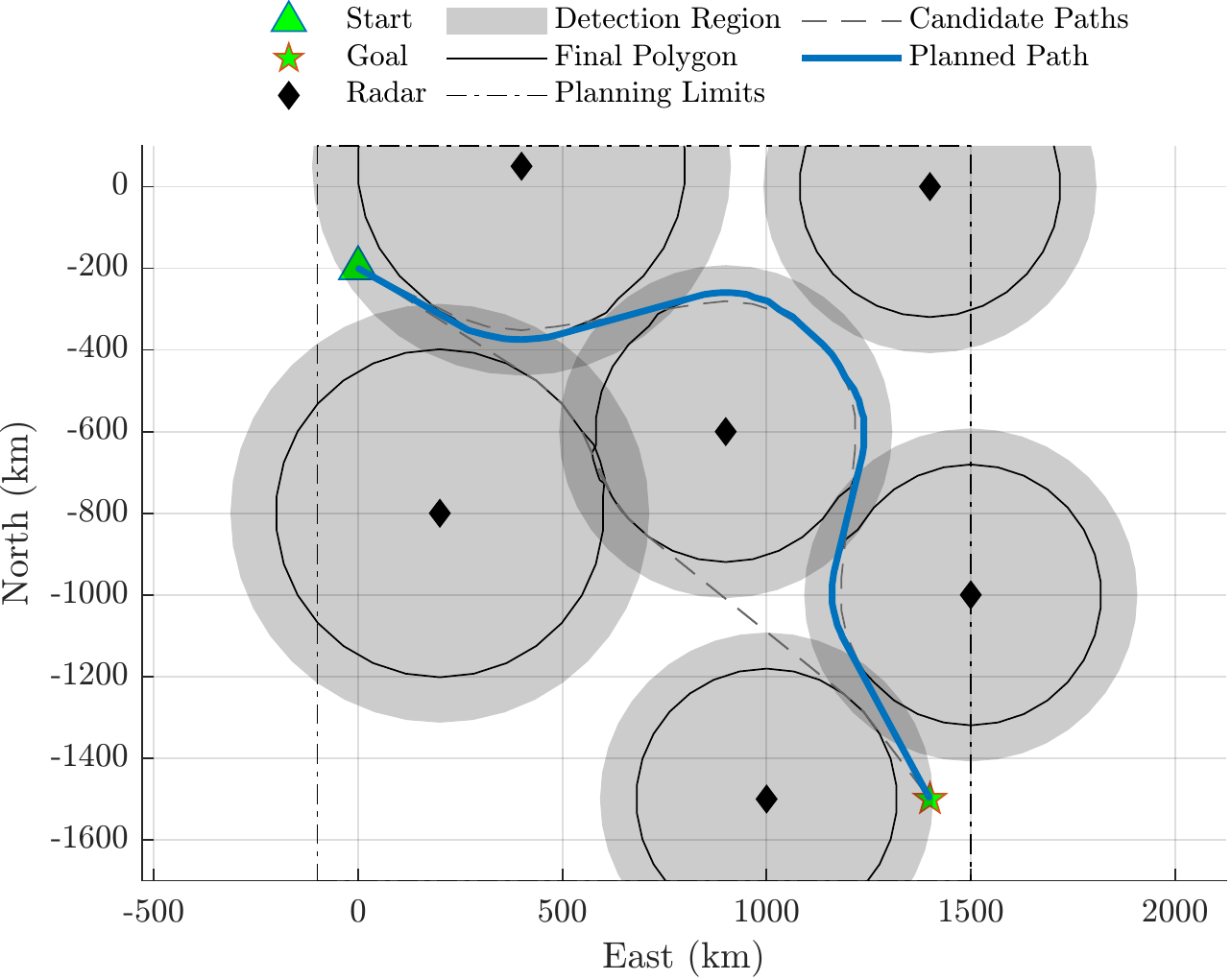}
		\par\end{centering}
	\caption{Visibility graph path planner results for Scenario 3 with six radar systems (diamonds), radar polygons (black lines), GPS-Denied regions (thick black rectangles), a candidate path (dashed gray line), and a planned path (blue line).\label{fig:res_scenario3}}
\end{figure}

\begin{figure}
	\begin{centering}
		\includegraphics[width=1\columnwidth]{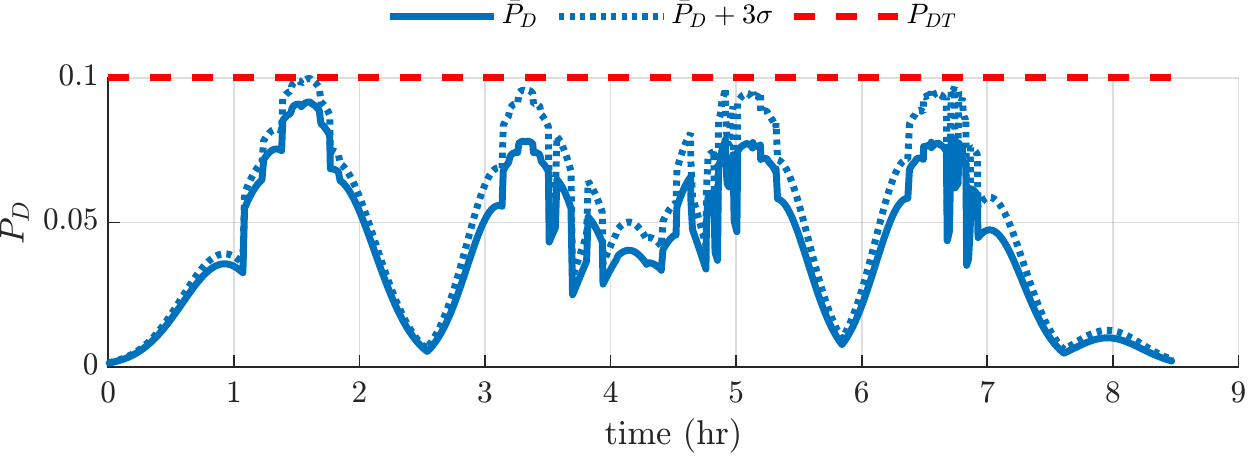}
		\par\end{centering}
	\caption{$P_D$ results for Scenario 3 with the $P_D$ threshold $P_{DT}=0.1$, nominal $\bar{P}_D$, and $\bar{P}_D + 3\sigma_{pd}$. The planner successfully finds a path through the radar detection region that maintains $\bar{P}_D + 3\sigma_{pd} < P_{DT}$. \label{fig:pd_res_scenario3}}
\end{figure}

\paperplan{Scenario 3}
{
    \begin{itemize}
		\item Navigation grade IMU
		\item Ellipsoid RCS, GPS-denied region, reduced radar constant variance
		\item 2D Plot showing path through radar detection regions
        \item PD Plot with $\bar{P}_D + 3\sigma_{pd}$
        \item Error budget combined with Scenario 2
    \end{itemize}
}
The third scenario illustrates the performance of the PDVG path planning algorithm in a radar detection region with several radar units.
For this scenario, the radar constant and radar uncertainty were lowered ($c_r=50$ and $20$, $\sigma_{pr}=100/3$, $\sigma_{cr}=1/3$) to fit more radar units in the planning region used for Scenarios 1 and 2.
Fig. \ref{fig:res_scenario3} shows the 2D map of the results for Scenario 3.
The dashed lines show the candidate paths considered by the PDVG path planner and the blue line shows the final planned path.
Fig. \ref{fig:pd_res_scenario3} shows the $P_D$ results for Scenario 3, which indicates the planner successfully finds a path through the radar detection region that maintains $\bar{P}_D + 3\sigma_{pd} < P_{DT}$. 
\revisionC{This scenario illustrates the utility of the PDVG path planner in complex detection environments that require expansions around several radar units.}

The results in this section have illustrated the methods presented in this paper in four ways.
First, the visibility graph path planner with the associated polygon adjustment policy is used to determine a feasible path that maintains $\bar{P}_D + 3\sigma_{pd} < P_{DT}$.
Second, the INS is modeled to obtain aircraft state uncertainty due to errors in measurement sources and GPS-Denied regions.
The resulting state covariance is used to inform the visibility graph path planner to \revisionC{find a path that maintains $P_D$ below the mission-specified threshold}.
Third, the LinCov models are used to generate error budgets to compare sources of uncertainty.
The error budget provided actionable information (i.e., improve IMU grade) which, when implemented, resulted in a shorter path between the radars that met the mission objectives.
Fourth, the PDVG planner described in this paper is applied to a congested detection region with several radar systems.

\section{Conclusion \label{sec:Conclusion}}
\paperplan{Summarize motivation}{}

Path planning for aircraft operating under threat of detection from ground-based radar systems must account for the probability of detection.
Several factors influence the detection risk of the aircraft including the aircraft pose, radar position, and radar performance characteristics.
In addition, uncertainty in each of these factors also influence the probability of detection.
Current radar detection path planning methods fail to consider the uncertainty in these factors when estimating the detection risk of a trajectory.
In practice, the uncertainty inherent in each of these factors is significant and influences the variability of the probability of detection.
\revisionC{In the scenarios shown in this paper, the variability of $P_D$ is a large fraction, up to $34.4\%$, of the nominal value.} 

\paperplan{Summarize approach}{}

This paper presents a method to propagate the covariance of the aircraft state and incorporate it with the uncertainty of the radar state into the radar detection model.
The method uses an inertial navigation system to propagate the covariance using IMU measurements and update the covariance with position, heading, and altitude measurements.
The nominal aircraft states and IMU measurements are generated using the ASG method from \cite{asg}.
The radar detection model is linearized about a nominal operating point and the variance of $P_D$ is calculated using the aircraft and radar state covariance matrices \cite{costley2022sensitivity,rds_ext}.

\paperplan{Summarize key findings}
{
    \begin{itemize}
        \item list
    \end{itemize}
}

These methods are used in the $P_D$ Visibility Graph (PDVG) path planner, where the nominal $P_D$ and standard deviation, $\sigma_{pd}$, are used to determine path validity and the polygon expansion policy.
This paper shows that the PDVG planner successfully plans paths that maintain $P_D + 3\sigma_{pd}$ below a threshold for three scenarios.
The error budgets generated by the path planner indicate that when the aircraft travels through GPS-Denied regions, measurement uncertainties in the IMU become the primary driver of the variance of $P_D$.
This variance is reduced by improving the IMU grade used in the aircraft model.

The results show that an advantage of using the methods presented in this paper is that the user can evaluate the sources of uncertainty and make actionable decisions.
For example, the IMU uncertainties were the primary driver of the variance of $P_D$ in Scenario 1, but when the IMU grade was improved, $\sigma_{pd}$ was reduced by $27.8\%$.
The improved IMU grade allowed the aircraft to travel between two radars which was infeasible with the original IMU grade.
\revisionC{The results in this paper show that uncertainty in the aircraft and radar states significantly impact the variability of $P_D$ and must be considered for path planning.
To address this, the PDVG planner is presented as a framework for incorporating the variability of $P_D$ in planning a path that maintains the probability of detection below a mission-specified threshold. 
 }

\bibliographystyle{IEEEtran}
\bibliography{full_bib}

\begin{thebibliography}{10}
\providecommand{\url}[1]{#1}
\csname url@samestyle\endcsname
\providecommand{\newblock}{\relax}
\providecommand{\bibinfo}[2]{#2}
\providecommand{\BIBentrySTDinterwordspacing}{\spaceskip=0pt\relax}
\providecommand{\BIBentryALTinterwordstretchfactor}{4}
\providecommand{\BIBentryALTinterwordspacing}{\spaceskip=\fontdimen2\font plus
\BIBentryALTinterwordstretchfactor\fontdimen3\font minus
  \fontdimen4\font\relax}
\providecommand{\BIBforeignlanguage}[2]{{%
\expandafter\ifx\csname l@#1\endcsname\relax
\typeout{** WARNING: IEEEtran.bst: No hyphenation pattern has been}%
\typeout{** loaded for the language `#1'. Using the pattern for}%
\typeout{** the default language instead.}%
\else
\language=\csname l@#1\endcsname
\fi
#2}}
\providecommand{\BIBdecl}{\relax}
\BIBdecl

\bibitem{Ceccarelli_micro_uav}
N.~Ceccarelli, J.~J. Enright, E.~Frazzoli, S.~J. Rasmussen, and C.~J.
  Schumacher, ``Micro uav path planning for reconnaissance in wind,'' in
  \emph{2007 American Control Conference}, 2007, pp. 5310--5315.

\bibitem{larson_path_nodate}
R.~Larson, M.~Pachter, and M.~Mears, ``Path {Planning} by {Unmanned} {Air}
  {Vehicles} for {Engaging} an {Integrated} {Radar} {Network},'' in
  \emph{{AIAA} {Guidance}, {Navigation}, and {Control} {Conference} and
  {Exhibit}}.\hskip 1em plus 0.5em minus 0.4em\relax American Institute of
  Aeronautics and Astronautics, 2001.

\bibitem{xiao-wei_path_2010}
F.~Xiao-wei, L.~Zhong, and G.~Xiao-guang, ``Path {Planning} for {UAV} in
  {Radar} {Network} {Area},'' in \emph{2010 {Second} {WRI} {Global} {Congress}
  on {Intelligent} {Systems}}, vol.~3, Dec. 2010, pp. 260--263.

\bibitem{kabamba_optimal_2012}
P.~T. Kabamba, S.~M. Meerkov, and F.~H.~Z. III,
  ``\BIBforeignlanguage{en}{Optimal {Path} {Planning} for {Unmanned} {Combat}
  {Aerial} {Vehicles} to {Defeat} {Radar} {Tracking}},''
  \emph{\BIBforeignlanguage{en}{Journal of Guidance, Control, and Dynamics}},
  May 2012.

\bibitem{marcum_statistical_1960}
J.~Marcum, ``A statistical theory of target detection by pulsed radar,''
  \emph{IRE Transactions on Information Theory}, vol.~6, no.~2, pp. 59--267,
  Apr. 1960.

\bibitem{swerling_probability_1954}
P.~Swerling, ``\BIBforeignlanguage{en}{Probability of {Detection} for
  {Fluctuating} {Targets}},'' RAND Corporation, Tech. Rep., Jan. 1954.

\bibitem{mahafza_matlab_2003}
B.~R. Mahafza and A.~Elsherbeni, \emph{\BIBforeignlanguage{en}{{MATLAB}
  {Simulations} for {Radar} {Systems} {Design}}}.\hskip 1em plus 0.5em minus
  0.4em\relax CRC Press, Dec. 2003.

\bibitem{mcfarland_motion_1999}
M.~McFarland, R.~Zachery, and B.~Taylor, ``Motion planning for reduced
  observability of autonomous aerial vehicles,'' in \emph{Proceedings of the
  1999 {IEEE} {International} {Conference} on {Control} {Applications} ({Cat}.
  {No}.{99CH36328})}, vol.~1, Aug. 1999, pp. 231--235 vol. 1.

\bibitem{bortoff_path_2000}
S.~Bortoff, ``Path planning for {UAVs},'' in \emph{Proceedings of the 2000
  {American} {Control} {Conference}. {ACC} ({IEEE} {Cat}. {No}.{00CH36334})},
  vol.~1, Jun. 2000, pp. 364--368 vol.1, iSSN: 0743-1619.

\bibitem{chandler_uav_2000}
P.~Chandler, S.~Rasmussen, and M.~Pachter, ``\BIBforeignlanguage{en}{{UAV}
  cooperative path planning},'' in \emph{\BIBforeignlanguage{en}{{AIAA}
  {Guidance}, {Navigation}, and {Control} {Conference} and {Exhibit}}}.\hskip
  1em plus 0.5em minus 0.4em\relax Dever,CO,U.S.A.: American Institute of
  Aeronautics and Astronautics, Aug. 2000.

\bibitem{pachter_optimal_2001}
M.~Pachter and J.~Hebert, ``Optimal aircraft trajectories for radar exposure
  minimization,'' in \emph{Proceedings of the 2001 {American} {Control}
  {Conference}. ({Cat}. {No}.{01CH37148})}, vol.~3, Jun. 2001, pp. 2365--2369
  vol.3, iSSN: 0743-1619.

\bibitem{moore_radar_2002}
F.~Moore, ``Radar cross-section reduction via route planning and intelligent
  control,'' \emph{IEEE Transactions on Control Systems Technology}, vol.~10,
  no.~5, pp. 696--700, Sep. 2002.

\bibitem{jun_path_2003}
M.~Jun and R.~D’Andrea, ``\BIBforeignlanguage{en}{Path {Planning} for
  {Unmanned} {Aerial} {Vehicles} in {Uncertain} and {Adversarial}
  {Environments}},'' in \emph{\BIBforeignlanguage{en}{Cooperative {Control}:
  {Models}, {Applications} and {Algorithms}}}, ser. Cooperative {Systems},
  S.~Butenko, R.~Murphey, and P.~M. Pardalos, Eds.\hskip 1em plus 0.5em minus
  0.4em\relax Boston, MA: Springer US, 2003, pp. 95--110.

\bibitem{costley2022sensitivity}
A.~Costley, R.~Christensen, G.~Droge, and R.~C. Leishman, ``Sensitivity of
  single-pulse radar detection to aircraft pose uncertainties,'' 2022.

\bibitem{savage_strapdown_2000}
P.~G. Savage, \emph{\BIBforeignlanguage{English}{Strapdown analytics}}.\hskip
  1em plus 0.5em minus 0.4em\relax Maple Plain, Minn.: Strapdown Associates,
  2000.

\bibitem{grewal_global_2020}
M.~S. Grewal, A.~P. Andrews, and C.~G. Bartone,
  \emph{\BIBforeignlanguage{en}{Global {Navigation} {Satellite} {Systems},
  {Inertial} {Navigation}, and {Integration}}}.\hskip 1em plus 0.5em minus
  0.4em\relax John Wiley \& Sons, Jan. 2020.

\bibitem{farrell_aided_2008}
J.~Farrell, \emph{Aided {Navigation}: {GPS} with {High} {Rate} {Sensors}},
  1st~ed.\hskip 1em plus 0.5em minus 0.4em\relax USA: McGraw-Hill, Inc., 2008.

\bibitem{maybeck_stochastic_1994}
P.~S. Maybeck, \emph{Stochastic {Models}, {Estimation}, and {Control}}.\hskip
  1em plus 0.5em minus 0.4em\relax New York: Navtech Book and Software Store,
  1994, vol.~1.

\bibitem{brown_kalman}
R.~G. {Brown} and P.~Y.~C. {Hwang}, \emph{{Introduction to random signals and
  applied Kalman filtering : with MATLAB exercises and solutions}}, 1997.

\bibitem{christensen_2014}
\BIBentryALTinterwordspacing
R.~S. Christensen and D.~Geller, ``Linear covariance techniques for closed-loop
  guidance navigation and control system design and analysis,''
  \emph{Proceedings of the Institution of Mechanical Engineers, Part G: Journal
  of Aerospace Engineering}, vol. 228, no.~1, pp. 44--65, 2014. [Online].
  Available: \url{https://doi.org/10.1177/0954410012467717}
\BIBentrySTDinterwordspacing

\bibitem{christensen2021closedloop}
\BIBentryALTinterwordspacing
R.~S. Christensen, G.~Droge, and R.~C. Leishman, ``Closed-{Loop} {Linear}
  {Covariance} {Framework} for {Path} {Planning} in {Static} {Uncertain}
  {Obstacle} {Fields},'' \emph{Journal of Guidance, Control, and Dynamics},
  vol.~45, no.~4, pp. 669--683, 2022, publisher: American Institute of
  Aeronautics and Astronautics \_eprint: https://doi.org/10.2514/1.G006228.
  [Online]. Available: \url{https://doi.org/10.2514/1.G006228}
\BIBentrySTDinterwordspacing

\bibitem{rds_ext}
\BIBentryALTinterwordspacing
A.~Costley, R.~Christensen, R.~C. Leishman, and G.~Droge, ``Sensitivity of
  single-pulse radar detection to radar state uncertainty,'' 2022. [Online].
  Available: \url{https://arxiv.org/abs/2203.11372}
\BIBentrySTDinterwordspacing

\bibitem{asg}
A.~Costley, R.~S. Christensen, R.~Leishman, and G.~Droge, ``Analytical aircraft
  state and imu signal generator from smoothed reference trajectory,''
  \emph{IEEE Transactions on Aerospace and Electronic Systems}, pp. 1--1, 2021.

\bibitem{north_analysis_1963}
D.~North, ``An {Analysis} of the factors which determine signal/noise
  discrimination in pulsed-carrier systems,'' \emph{Proceedings of the IEEE},
  vol.~51, no.~7, pp. 1016--1027, Jul. 1963.

\bibitem{beard_randy_small_2012}
{Beard, Randy} and {McLain, Timothy}, \emph{Small {Unmanned} {Aircraft}
  {Theory} and {Practice}}, 2012.

\bibitem{zanetti_rotations_2019}
R.~Zanetti, ``\BIBforeignlanguage{en}{Rotations, {Transformations}, {Left}
  {Quaternions}, {Right} {Quaternions}?}'' \emph{\BIBforeignlanguage{en}{The
  Journal of the Astronautical Sciences}}, vol.~66, no.~3, pp. 361--381, Sep.
  2019.

\bibitem{carpenter_navigation_2018}
J.~R. Carpenter and C.~N. D’Souza, ``\BIBforeignlanguage{en}{Navigation
  {Filter} {Best} {Practices}},'' National Aeronautics and Space
  Administration, {NASA} {Technical} {Report}, Apr. 2018.

\bibitem{lear_kalman_1985}
W.~Lear, ``Kalman {Filtering} {Techniques},'' NASA Johnson Space Center, {NASA}
  {Technical} {Report} JSC-20688, Sep. 1985.

\bibitem{scheuer_continuous-curvature_1997}
A.~Scheuer and T.~Fraichard, ``Continuous-curvature path planning for car-like
  vehicles,'' in \emph{Proceedings of the 1997 {IEEE}/{RSJ} {International}
  {Conference} on {Intelligent} {Robot} and {Systems}. {Innovative} {Robotics}
  for {Real}-{World} {Applications}. {IROS} 97}, vol.~2, Sep. 1997, pp.
  997--1003 vol.2.

\bibitem{dijkstra_note_1959}
E.~Dijkstra, ``A {Note} on {Two} {Problems} in {Connexion} with {Graphs},''
  \emph{Numerische mathematik}, vol.~1, no.~1, pp. 269--271, 1959.

\bibitem{hart_astar}
P.~E. Hart, N.~J. Nilsson, and B.~Raphael, ``A formal basis for the heuristic
  determination of minimum cost paths,'' \emph{IEEE Transactions on Systems
  Science and Cybernetics}, vol.~4, no.~2, pp. 100--107, 1968.

\bibitem{blanco2010tutorial}
J.-L. Blanco, ``A tutorial on se (3) transformation parameterizations and
  on-manifold optimization,'' \emph{University of Malaga, Tech. Rep}, vol.~3,
  p.~6, 2010.

\end{thebibliography}

\appendices{}
% \section{RCS Azimuth and Elevation Angle Derivation \label{app:az_el}}
% \input{sections/Appendices.tex}

% \section{Radar Model Jacobians \label{app:rad_jacs}}
% \input{sections/AP_RadarJacobians.tex}

\section{Quaternion to Euler Angle Linearization \label{app:quat2eul}}
This appendix derives the linearization of the quaternion to Euler angle attitude representations with a small angle assumption.
The linearized model is used to define the matrix $M_a$ which transforms the aircraft state of the INS model $\boldsymbol{\hat{x}}$ defined in \eqref{eq:nav_model} to the aircraft state of the radar model $\boldsymbol{x_a}$ given by $\boldsymbol{x_a} = \begin{bmatrix} \boldsymbol{p_a^n} & \boldsymbol{\Theta_a} \end{bmatrix}^\intercal$.

The general quaternion to Euler transformation \cite{blanco2010tutorial} is given by
\begin{equation}
    \begin{bmatrix} \phi \\ \theta \\ \psi \end{bmatrix} = \begin{bmatrix} \arctan{\frac{2\left(q_0 q_1 + q_2 q_3\right)}{1-2\left(q_1^2 + q_2^2\right)}} \\ \arcsin{2(q_0q_2-q_3q_1)} \\ \arctan{\frac{2\left(q_0 q_3 + q_1 q_2\right)}{1-2\left(q_2^2 + q_3^2\right)}} \end{bmatrix}
\end{equation}
The first element of the error quaterion defined in \eqref{eq:quat_error} is 1 so let $q_0 = 1$ then
\begin{equation}
    \begin{bmatrix} \phi \\ \theta \\ \psi \end{bmatrix} = \begin{bmatrix} \arctan{\frac{2\left(q_1 + q_2 q_3\right)}{1-2\left(q_1^2 + q_2^2\right)}} \\ \arcsin{2(q_2-q_3q_1)} \\ \arctan{\frac{2\left(q_3 + q_1 q_2\right)}{1-2\left(q_2^2 + q_3^2\right)}} \end{bmatrix} \label{eq:quat2eul}
\end{equation}

The Jacobian of \eqref{eq:quat2eul} with respect to $\boldsymbol{q}_{1:3}$ is $\partiald{\boldsymbol{\Theta}}{\boldsymbol{q}_{1:3}}$.
The elements of $\partiald{\boldsymbol{\Theta}}{\boldsymbol{q}_{1:3}}$ are given by
\begin{align}
    \partiald{\phi}{q_1} &= -\frac{\frac{2}{q_a} - \frac{q_1(2q_1+2q_2q_3)4}{q_a^2}q_a^2}{q_a^2 + (2q_1 + 2q_2 q_3)} \\
    \partiald{\phi}{q_2} &= -\frac{\frac{2q_3}{q_a} - \frac{q_2(2q_1+2q_2q_3)4}{q_a^2}q_a^2}{q_a^2 + (2q_1 + 2q_2 q_3)} \\ 
    \partiald{\phi}{q_3} &= -\frac{2q_1^2 + 2q_2^2 -1}{(2q_1^2 + 2q_2^2 -1)^2 + (2q_1 + 2q_2 q_3)^2} \\
    \partiald{\theta}{q_1} &= -\frac{2q_3}{\sqrt{1-(2q_2-2q_1q_3)^2}} \\
    \partiald{\theta}{q_2} &= \frac{2}{\sqrt{1-(2q_2-2q_1q_3)^2}} \\
    \partiald{\theta}{q_3} &=  -\frac{2q_1}{\sqrt{1-(2q_2-2q_1q_3)^2}} \\ 
    \partiald{\psi}{q_1} &= -\frac{2q_2q_b}{q_b^2 + (2q_3+2q_1q_2)^2} \\
    \partiald{\psi}{q_2} &= -\frac{\frac{2q_1}{q_b} - \frac{q_2(2q_3+2q_1q_2)4}{q_b^2}q_b^2}{q_b^2 + (2q_3 + 2q_1 q_2)^2} \\
    \partiald{\psi}{q_3} &= -\frac{\frac{2}{q_b} - \frac{q_3(2q_3+2q_1q_2)4}{q_b}q_b^2}{q_b^2 + (2q_3 + 2q_1 q_2)}          
\end{align}
where
\begin{align}
    q_a &= 2q_1^2+2q_2^2-1 \\
    q_b &= 2q_2^2+2q_3^2-1.
\end{align}
Evaluating the partial derivatives at the nominal $q_1=q_2=q_3=0$ yields the final Jacobian of the quaternion to Euler angle transformation as
\begin{equation}
    \partiald{\boldsymbol{\Theta}}{\boldsymbol{q}_{1:3}} \Bigg\rvert_{\boldsymbol{\delta \bar{q}}} = 2\boldsymbol{I}_{3\times3}. \label{eq:quat2eul_jacobian}
\end{equation}

% Let
% \begin{equation}
%     J_\Theta = \partiald{\boldsymbol{\Theta}}{\boldsymbol{\theta}} \Bigg\rvert_{\boldsymbol{\bar{\theta}}}
% \end{equation}
% then the linearized model can be used to transform the gaussian random variable as
% \begin{equation}
%     \boldsymbol{\Theta} = J_\Theta \boldsymbol{\theta}
% \end{equation}
% and the covariance is transformed with
% \begin{equation}
%     C_{\boldsymbol{\Theta}\boldsymbol{\Theta}} = J_\Theta C_{\theta \theta} J_\Theta^\intercal.
% \end{equation}

The quaterion to Euler angle Jacobian defined in \eqref{eq:quat2eul_jacobian} is used to transform the INS model state $\boldsymbol{\hat{x}}$ as defined in \eqref{eq:nav_model} to the aircraft state in the radar model $\boldsymbol{x_a}$ as  
\begin{align}
    \boldsymbol{x_a} &= \begin{bmatrix} I_{3 \times 3} & 0_{3 \times 3} & 0_{3 \times 3} & 0_{3 \times 3} & 0_{3 \times 3} \\
                            0_{3 \times 3} & 0_{3 \times 3} & 2I_{3 \times 3} & 0_{3 \times 3} & 0_{3 \times 3}                       
                        \end{bmatrix} \boldsymbol{x} \\
                     &= M_a \boldsymbol{x}.
\end{align}

\section{Covariance Propagation using Lear's Method \label{app:lears}}
While the error state model in \eqref{eq:cov-err-state-diffeq} is an accurate representation of the error dynamics, the error states are more efficiently propagated using the state transition matrix (STM).
This is represented in discrete time as
\begin{equation}
\delta\boldsymbol{x}_{k}=\Phi\left(t_{k},t_{k-1}\right)\delta\boldsymbol{x}_{k-1}+\boldsymbol{w}_{k-1}
\end{equation}
where $\Phi\left(t_{k},t_{k-1}\right)$ is the STM from the $t_{k-1}$
to $t_{k}$ , and $\boldsymbol{w}_{k-1}$ is the integrated process
noise over the same time interval given by
\begin{equation}
\boldsymbol{w}_{k-1}=\int_{t_{k-1}}^{t_{k}}\Phi\left(t_{k},\tau\right)BQ\left(\tau\right)\boldsymbol{w}\left(\tau\right)d\tau.
\end{equation}

The covariance of the estimation errors is also propagated using the STM using
\begin{equation}
P_{k}^{-}=\Phi\left(t_{k},t_{k-1}\right)P_{k-1}^{-}\Phi^{T}\left(t_{k},t_{k-1}\right)+Q_{k-1}
\end{equation}
where $Q_{k-1}$ is the covariance of the integrated process noise defined by
\begin{align}
Q_{k-1} &= E\left[\boldsymbol{w}_{k-1}\boldsymbol{w}_{k-1}^{T}\right]\nonumber \\
 &= \int_{t_{k-1}}^{t_{k}}\Phi\left(t_{k},\tau\right)B\left(\tau\right)Q\left(\tau\right)B^{T}\left(\tau\right)\Phi^{T}\left(t_{k},\tau\right)d\tau. \label{eq:cov-int-proc-noise}
\end{align}
The following paragraphs provide a derivation for $\Phi\left(t_{k},t_{k-1}\right)$ and $Q_{k-1}$. 
The STM is defined as the matrix which satisfies the differential equation \cite{maybeck_stochastic_1994}
\begin{equation}
\dot{\Phi}\left(t_{k+1},t_{k}\right)=F\left(t\right)\Phi\left(t_{k+1},t_{k}\right) \label{eq:cov-stmdot}
\end{equation}
where $\Phi\left(t_{k},t_{k}\right)=I_{n\times n}$.
While it is possible to numerically integrate \eqref{eq:cov-stmdot},
approximation of the integral over small time frames is sufficient
for many practical filters and is much more efficient. There are many
good approximations for this purpose \cite{carpenter_navigation_2018},
however Lear's method is chosen for this application \cite{lear_kalman_1985}.
Lear's method for approximating the state transition matrix is given
by
\begin{equation}
\Phi\left(t_{k+1},t_{k}\right)\approx I+\frac{\Delta t}{2}\left(F_{k}+F_{k-1}\right)+\frac{\Delta t^{2}}{2}F_{k}F_{k-1}\label{eq:cov-lear}
\end{equation}
where $F_{k}=F\left(t_{k}\right)$ and $\Delta t=t_{k}-t_{k-1}$.
To aid in the derivation that follows, the error states defined in \eqref{eq:error_states} are partitioned into vehicle states and parameter states as
\begin{equation}
\delta\boldsymbol{x}=\left[\begin{array}{cc}
\delta\boldsymbol{x}_{v} & \delta\boldsymbol{x}_{p}\end{array}\right]^{T}
\end{equation}
where the vehicle states comprise the position, velocity, and attitude
errors
\begin{equation}
\delta\boldsymbol{x}_{v}=\left[\begin{array}{ccc}
\delta\boldsymbol{p}^{n} & \delta\boldsymbol{v}^{n} & \delta\boldsymbol{\theta}{}_{b}^{n}\end{array}\right]^{T}
\end{equation}
and the parameter states include the accelerometer and gyro biases.
\begin{equation}
\delta\boldsymbol{x}_{p}=\left[\begin{array}{cc}
\delta\boldsymbol{b}_{a}^{b} & \delta\boldsymbol{b}_{g}^{b}\end{array}\right]^{T}
\end{equation}
The state coupling matrix $F_k$ has the same form as \eqref{eq:F_hat} and is partitioned as 
\begin{equation}
F_{k}=\left[\begin{array}{cc}
F_{vv} & F_{vp}\\
0_{6\times9} & F_{pp}
\end{array}\right]
\end{equation}
and the STM is partitioned as
\begin{equation}
\Phi\left(t_{k+1},t_{k}\right)=\Phi_{k}=\left[\begin{array}{cc}
\Phi_{vv} & \Phi_{vp}\\
\Phi_{pv} & \Phi_{pp}
\end{array}\right]. \label{eq:stm_partition}
\end{equation}

Since the parameter states are modeled as FOGM processes, independent
of the vehicle states, the corresponding elements of the STM are known
analytically as
\begin{equation}
\Phi_{pv}=0_{6\times9}
\end{equation}
and
\begin{equation}
\Phi_{pp}=\left[\begin{array}{cc}
\exp\left(-\Delta t/\tau_{a}\right)I_{3\times3} & 0_{3\times3}\\
0_{3\times3} & \exp\left(-\Delta t/\tau_{g}\right)I_{3\times3}
\end{array}\right]\label{eq:cov-param-stm}
\end{equation}
The upper-left element of the STM from \eqref{eq:stm_partition} is approximated using \eqref{eq:cov-lear} to yield
\begin{equation}
    \Phi_{vv}\approx\left[\begin{array}{ccc}
    I_{3\times3} & \Delta tI_{3\times3} & \frac{\Delta t^{2}}{2}\left[\hat{\boldsymbol{\nu}}^n_{k-1}\right]\times\\
    0_{3\times3} & I_{3\times3} & \frac{\Delta t}{2}\left[\hat{\boldsymbol{\nu}}^n_{\Sigma}\right]\times\\
    0_{3\times3} & 0_{3\times3} & I_{3\times3}
    \end{array}\right]\label{eq:cov-vehicle-stm}
\end{equation}
where
\begin{eqnarray}
    \hat{\boldsymbol{\nu}}^n_{k-1} &=& \hat{T}_{b,k-1}^{n}\left(\tilde{\boldsymbol{\nu}}_{k-1}^{b}-\hat{\boldsymbol{b}}_{a,k-1}^{b}\right) \\
    \hat{\boldsymbol{\nu}}^n_{k} &=& \hat{T}_{b,k-1}^{n}\left(\tilde{\boldsymbol{\nu}}_{k}^{b}-\hat{\boldsymbol{b}}_{a,k}^{b}\right)
\end{eqnarray}
and 
\begin{equation}
    \hat{\boldsymbol{\nu}}^n_{\Sigma} = \hat{\boldsymbol{\nu}}^n_{k} + \hat{\boldsymbol{\nu}}^n_{k-1}.
\end{equation}

The lower-left element of the STM from \eqref{eq:stm_partition} is approximated using \eqref{eq:cov-lear} to yield
\begin{equation}
\Phi_{vp}\approx\left[\begin{array}{cc}
-\frac{\Delta t^{2}}{2}\hat{T}_{b,k-1}^{n} & 0_{3\times3}\\
-\frac{\Delta t}{2}\hat{T}_{b,\Sigma}^{n}+\frac{\Delta t^{2}}{2\tau_{a}}\hat{T}_{b,k}^{n} & \frac{\Delta t^{2}}{2}\left\{ \left[\hat{\boldsymbol{\nu}}^n_{k}\right]\times\right\} \hat{T}_{b,k-1}^{n}\\
0_{3\times3} & \frac{\Delta t}{2}\hat{T}_{b,\Sigma}^{n}-\frac{\Delta t^{2}}{2\tau_{g}}\hat{T}_{b,k}^{n}
\end{array}\right]\label{eq:cov-vehicle-parame-stm}
\end{equation}
where
\begin{eqnarray}
\hat{T}_{b,\Sigma}^{n} &=& \hat{T}_{b,k}^{n}+\hat{T}_{b,k-1}^{n}.
\end{eqnarray}

The integrated process noise $Q_{k-1}$ in \eqref{eq:cov-int-proc-noise} is derived in the following paragraphs. 
Begin by partitioning the dynamics coupling matrix and noise density matrix
\begin{equation}
B=\left[\begin{array}{cc}
B_{v,n} & 0_{9\times6}\\
0_{6\times6} & I_{6\times6}
\end{array}\right]
\end{equation}

\begin{equation}
Q=\left[\begin{array}{cc}
Q_{n} & 0_{6\times6}\\
0_{6\times6} & Q_{w}
\end{array}\right]
\end{equation}
where 
\begin{equation}
B_{v,n}=\left[\begin{array}{cc}
0_{3\times3} & 0_{3\times3}\\
-\hat{T}_{b}^{n} & 0_{3\times3}\\
0_{3\times3} & \hat{T}_{b}^{n}
\end{array}\right]\label{eq:cov-Bvn}
\end{equation}
\begin{equation}
Q_{n}=\left[\begin{array}{cc}
Q_{\nu} & 0_{3\times3}\\
0_{3\times3} & Q_{\omega}
\end{array}\right]\label{eq:cov-Qn}
\end{equation}
\begin{equation}
Q_{w}=\left[\begin{array}{cc}
q_{a}I_{3\times3} & 0_{3\times3}\\
0_{3\times3} & q_{g}I_{3\times3}
\end{array}\right]\label{eq:cov-param-psd}
\end{equation}

The integral of \eqref{eq:cov-int-proc-noise} can be expressed as
\begin{equation}
Q_{k-1}=\left[\begin{array}{cc}
Q_{vv} & Q_{vp}\\
Q_{pv} & Q_{pp}
\end{array}\right]
\end{equation}
where 
\begin{eqnarray}
Q_{vv} & = & \int_{t_{k-1}}^{t_{k}}\Phi_{vv}\left(t_{k},\tau\right)B_{v,n}\left(\tau\right)Q_{n}B_{v,n}^{T}\left(\tau\right)\Phi_{vv}^{T}\left(t_{k},\tau\right)\nonumber \\
 &  & +\Phi_{vp}\left(t_{k},\tau\right)Q_{w}\Phi_{vp}^{T}\left(t_{k},\tau\right)d\tau\label{eq:cov-vehicle-Q} \\
Q_{vp} &=& Q_{pv}^{T}=\int_{t_{k-1}}^{t_{k}}\Phi_{vp}\left(t_{k},\tau\right)Q_{w}\Phi_{pp}^{T}\left(t_{k},\tau\right)d\tau\label{eq:cov-vp-Q}
\end{eqnarray}
and
\begin{equation}
Q_{pp}=\int_{t_{k-1}}^{t_{k}}\Phi_{pp}\left(t_{k},\tau\right)Q_{w}\Phi_{pp}^{T}\left(t_{k},\tau\right)d\tau\label{eq:cov-param-Q}
\end{equation}

The integrals in \eqref{eq:cov-vehicle-Q}-\eqref{eq:cov-param-Q} are derived separately in the following paragraphs. 
The integrated process noise for the parameter states is approximated as%
\begin{equation}
Q_{pp}\approx\left[\begin{array}{cc}
q_{a} \frac{\tau_a}{2} e_a I_{3\times3} & 0_{3\times3}\\
0_{3\times3} & q_{g}\frac{\tau_g}{2} e_g I_{3\times3}
\end{array}\right].
\end{equation}
where
\begin{eqnarray}
    e_a &=& 1-\exp(\frac{-2 \Delta t}{\tau_a}) \\
    e_g &=& 1-\exp(\frac{-2 \Delta t}{\tau_g}).
\end{eqnarray}

The integrated process noise of the coupled parameter and vehicle
states can be derived as%
\begin{eqnarray}
Q_{vp}&=&\left[\begin{array}{c}
-q_{a}\hat{T}_{b,k-1}^{n}c_{a} \\ 
-q_{a}\hat{T}_{b,\Sigma}^{n}c_t+q_{a}\hat{T}_{b,k}^{n}\frac{c_{a}}{\tau_{a}}  \\
0_{3\times3}
\end{array}\right. \nonumber \\
& & \qquad \left.\begin{array}{c}
0_{3\times3}\\
q_{g}\left\{ \left[\hat{\boldsymbol{\nu}}^n_{k}\right]\times\right\} \hat{T}_{b,k-1}^{n}c_{a}\\
q_{g}\hat{T}_{b,\Sigma}^{n}c_t-q_{g}\hat{T}_{b,k}^{n}c_{a} \\
\end{array}\right]
\end{eqnarray}
where
\begin{eqnarray}
c_{a} &=& \tau_{a}^{3} - \frac{1}{2}\tau_{a}\left(2\tau_{a}^{2}+2\tau_{a}\Delta t+\Delta t^{2}\right)\exp(\frac{\Delta t}{\tau_a}) \\
% c_{g} &=& \tau_{g}^{3}+\frac{1}{2}\left(\Delta t-\tau_{g}\right)\left(2\tau_{g}^{2}+2\tau_{g}\Delta t+\Delta t^{2}\right) \\
c_{t} &=& \frac{\tau_a^2}{2} - \frac{\tau_a}{2}(\tau_a+\Delta t)\exp(\frac{-\Delta t}{\tau_a})
\end{eqnarray}

The integrated process noise of the vehicle states is partitioned as%
\begin{equation}
Q_{vv}=\left[\begin{array}{ccc}
Q_{vv,11} & Q_{vv,12} & Q_{vv,13}\\
Q_{vv,21} & Q_{vv,22} & Q_{vv,23}\\
Q_{vv,31} & Q_{vv,32} & Q_{vv,33}
\end{array}\right]\label{eq:integratedProcessNoise}
\end{equation}
where the equations for each entry are given by

\begin{align}
Q_{vv,11} &= \frac{\Delta t^{3}}{3}\mathcal{Q}_{\nu,k} \nonumber \\
 & +\frac{\Delta t^{5}}{20}\left(\mathcal{Q}_{b,k-1}^{n}+q_{a}\hat{T}_{b,k-1}^{n}\left(\hat{T}_{b,k-1}^{n}\right)^{T}\right)
\end{align}

\begin{align}
Q_{vv,12} &= \frac{\Delta t^{2}}{2}\mathcal{Q}_{\nu,k} \nonumber \\
 & +\frac{\Delta t^{4}}{16}\left(\left\{ \left[\hat{\boldsymbol{\nu}}^n_{k-1}\right]\times\right\} \mathcal{Q}_{\omega,k}\left\{ \left[\hat{\boldsymbol{\nu}}^n_{\Sigma}\right]\times\right\} ^{T}\right. \nonumber \\
 & \left.+q_{a}\hat{T}_{b,k-1}^{n}\hat{T}_{b,\Sigma}^{n}{}^{T}\right)-q_{a}\frac{\Delta t^{5}}{20\tau_{a}}\hat{T}_{b,k-1}^{n}\left(\hat{T}_{b,k}^{n}\right)^{T}
\end{align}

\begin{align}
Q_{vv,13} &= \frac{1}{6}\Delta t^{3}\left\{ \left[\hat{\boldsymbol{\nu}}^n_{k-1}\right]\times\right\} \mathcal{Q}_{\nu,k}
\end{align}

\begin{align}
Q_{vv,21} &= \frac{\Delta t^{2}}{2}\mathcal{Q}_{\nu,k} \nonumber \\
 & +\frac{\Delta t^{4}}{16}\left(\left\{ \left[\hat{\boldsymbol{\nu}}^n_{\Sigma}\right]\times\right\} \mathcal{Q}_{\omega,k}\left\{ \left[\hat{\boldsymbol{\nu}}^n_{k-1}\right]\times\right\} ^{T}\right. \nonumber \\
 & \left.+q_{a}\hat{T}_{b,\Sigma}^{n}\left(\hat{T}_{b,k-1}^{n}\right)^{T}\right)-q_{a}\frac{\Delta t^{5}}{20\tau_{a}}\hat{T}_{b,k-1}^{n}\left(\hat{T}_{b,k}^{n}\right)^{T}
\end{align}

\begin{align}
Q_{vv,22} &= \Delta t\mathcal{Q}_{\nu,k}+\frac{\Delta t^{3}}{12}\left(\left\{ \left[\hat{\boldsymbol{\nu}}^n_{\Sigma}\right]\times\right\} \mathcal{Q}_{\omega,k}\left\{ \left[\hat{\boldsymbol{\nu}}^n_{\Sigma}\right]\times\right\} ^{T}\right. \nonumber \\
 & \left.+q_{a}\hat{T}_{b,\Sigma}^{n}\left(\hat{T}_{b,\Sigma}^{n}\right)^{T}\right) \nonumber \\
 &  -q_{a}\frac{\Delta t^{4}}{16\tau_{a}}\hat{T}_{b,k}^{n}\left[\hat{T}_{b,k}^{n}\left(\mathcal{\hat{T}}_{b,\Sigma}^{n}\right)^{T}+\mathcal{\hat{T}}_{b,\Sigma}^{n}\left(\hat{T}_{b,k}^{n}\right)^{T}\right] \nonumber \\
 & +\frac{\Delta t^{5}}{20}\left(\frac{q_{a}}{\tau_{a}^{2}}\hat{R}_{b,k}^{n}\left(\hat{R}_{b,k}^{n}\right)^{T}\right. \nonumber \\
 & \left.+q_{g}\left\{ \left[\hat{\boldsymbol{\nu}}^n_{k}\right]\times\right\} \hat{T}_{b,k-1}^{n}\left(\hat{T}_{b,k-1}^{n}\right)^{T}\left\{ \left[\hat{\boldsymbol{\nu}}^n_{k}\right]\times\right\} ^{T}\right)
\end{align}

\begin{align}
Q_{vv,23} &= \frac{\Delta t^{2}}{4}\left\{ \left[\hat{\boldsymbol{\nu}}^n_{\Sigma}\right]\times\right\} \mathcal{Q}_{\omega,k}\nonumber \\
 & +q_{g}\frac{\Delta t^{4}}{16}\left\{ \left[\hat{\boldsymbol{\nu}}^n_{k}\right]\times\right\} \hat{T}_{b,k-1}^{n}\left(\hat{T}_{b,\Sigma}^{n}\right)^{T} \nonumber \\
 & -q_{g}\frac{\Delta t^{5}}{20\tau_{g}}\left\{ \left[\hat{\boldsymbol{\nu}}^n_{k}\right]\times\right\} \hat{T}_{b,k-1}^{n}\left(\hat{T}_{b,k}^{n}\right)^{T}
\end{align}

\begin{align}
Q_{vv,31} &= \frac{\Delta t^{3}}{6}\mathcal{Q}_{\omega,k}\left\{ \left[\hat{\boldsymbol{\nu}}^n_{k-1}\right]\times\right\} ^{T}
\end{align}

\begin{align}
Q_{vv,32} &= \frac{\Delta t^{2}}{4}\mathcal{Q}_{\omega,k}\left\{ \left[\hat{\boldsymbol{\nu}}^n_{\Sigma}\right]\times\right\} ^{T} \nonumber \\
 & +q_{g}\frac{\Delta t^{4}}{16}\hat{T}_{b,\Sigma}^{n}\left(\left\{ \left[\hat{\boldsymbol{\nu}}^n_{k}\right]\times\right\} \hat{T}_{b,k-1}^{n}\right)^{T} \nonumber \\
 & -q_{g}\frac{\Delta t^{5}}{20\tau_{g}}\hat{T}_{b,k}^{n}\left(\left\{ \left[\hat{\boldsymbol{\nu}}^n_{k}\right]\times\right\} \hat{T}_{b,k-1}^{n}\right)^{T}
\end{align}

\begin{align}
Q_{vv,33} &= \Delta t\mathcal{Q}_{\omega,k}+q_{g}\frac{\Delta t^{3}}{12}\hat{T}_{b,\Sigma}^{n}\left(\hat{T}_{b,\Sigma}^{n}\right)^{T} \nonumber \\
 & -q_{g}\frac{\Delta t^{4}}{16\tau_{g}}\left[\hat{T}_{b,k}^{n}\left(\hat{T}_{b,\Sigma}^{n}\right)^{T}+\hat{T}_{b,\Sigma}^{n}\left(\hat{T}_{b,k}^{n}\right)^{T}\right] \nonumber \\
 & +q_{g}\frac{\Delta t^{5}}{20\tau_{g}^{2}}\hat{T}_{b,k}^{n}\left(\hat{T}_{b,k}^{n}\right)^{T}
\end{align}
and
\begin{align}
    \mathcal{Q}_{b,k-1}^{n} &= \left\{ \left[\hat{\boldsymbol{\nu}}^n_{k-1}\right]\times\right\} \hat{T}_{b,k}^{n}Q_{\omega}\left(\hat{T}_{b,k}^{n}\right)^{T}\left\{ \left[\hat{\boldsymbol{\nu}}^n_{k-1}\right]\times\right\} ^{T}
\end{align}
\begin{align}
    \mathcal{Q}_{\nu,k} &= \hat{T}_{b,k}^{n}Q_{\nu}\left(\hat{T}_{b,k}^{n}\right)^{T}
\end{align}
\begin{align}
    \mathcal{Q}_{\omega,k} &= \hat{T}_{b,k}^{n}Q_{\omega}\left(\hat{T}_{b,k}^{n}\right)^{T}
\end{align}
\begin{align}
    \hat{R}_{b,\Sigma}^{n} &= \hat{T}_{b,k-1}^{n}+\hat{T}{}_{b,k}^{n}.
\end{align}

\end{document}